\documentclass[12pt,a4paper]{article}

\usepackage{color,amsmath,amsfonts,amssymb,upref,mathrsfs,bm}
\usepackage{graphicx}
\usepackage{slashed}
\usepackage{epsfig}
\usepackage[english]{babel}
\usepackage{color}
\def\({\left(}
\def\){\right)}
\def\[{\left[}
\def\]{\right]}
\def\be{\begin{equation}}
\def\ee{\end{equation}}
\def\a{\alpha}
\def\b{\beta}
\def\g{\gamma}

\def\s{\sigma}

\def\sh{\sqrt\h}
\def\t{\tau}
\def\vt{\vartheta}
\def\h{\hbar}

\def\hb{\hat\beta}
\def\vb{\check\beta}

\def\hvt{\hat\vartheta}
\def\q{{\bf q}}
\def\g{{\bf g}}
\def\f{{\bf f}}

\def\ps{{\bm \psi}}
\def\ph{{\bm \phi}}
\def\vph{{\bm \varphi}}
\def\vvph{\hat{\bm \varphi}}
\def\cvph{\check{\bm \varphi}}

\def\brr{}

\def\co{k}
\def\n{\overline n}
\def\vtn#1#2{\vartheta_{#2}^{(#1)}}
\def\hvtn#1#2{\hat\vartheta_{#2}^{(#1)}}
\def\psn#1{{\ps^{(#1)}}}
\def\phn#1{{\ph^{(#1)}}}

\def\vvphn#1{{\vvph^{(#1)}}}
\def\cvphn#1{{\cvph^{(#1)}}}

\def\hbn#1{{\hb^{(#1)}}}
\def\vbn#1{{\vb^{(#1)}}}

\def\cn#1#2{{c_{#2}^{(#1)}}}
\def\hcn#1#2{{\hat{c}_{#2}^{(#1)}}}
\def\I{{\bm{ \mathcal I}}}
\def\P{{\bm{ \mathcal P}}}
\def\G{{\bm \Gamma}}
\def\Ps{{\bm \Psi}}
\def\Ph{{\bm \Phi}}
\def\H{{\bm{ \mathcal H}}}

\def\K{{\bm{ \mathcal K}}}
\def\hK{\hat{\bm{ \mathcal K}}}
\def\Kc{{ \mathcal K}}
\def\B{\bm{ \mathcal B}}
\def\Bc{{ \mathcal B}}

\def\T{\bm{ \mathcal T}}
\def\Tc{{ \mathcal T}}

\def\hPs{\hat{\bm \Psi}}
\def\hPh{\hat{\bm \Phi}}
\def\Om{{\bm \Omega}}

\def\Phn#1#2{{\Ph_{#2}^{(#1)}}}
\def\hPhn#1#2{{\hat\Ph_{#2}^{(#1)}}}
\def\Bn#1{\B^{(#1)}}
\def\Bcn#1{\Bc^{(#1)}}
\def\Kcn#1{\Kc^{(#1)}}
\def\Kn#1{\K^{(#1)}}
\def\Kn#1{\K^{(#1)}}
\def\Kcn#1{\Kc^{(#1)}}

\def\wt{\widetilde}

\def\Ref#1{(\ref{#1})}
\def\={ \mathop{=}}
\def\seq{ \mathop{\simeq}}
\def\arg{ \mathop{\rm arg}}
\def\Im{ \mathop{\rm Im}}
\def\Re#1{ \mathop{\rm Re}\!#1 \,}
\def\exp#1{ \mathop{\rm exp}\nolimits\hskip-.7mm\left\{#1\right\}}
\def\sgn{ \mathop{\rm sgn}}

\def\part#1{}

\begin{document}

\title{
A unifying asymptotic approach for nonadiabatic transitions near pairs of real  or complex turning points\\
}

\author{Ignat Fialkovsky
\footnote{CMCC-Universidade Federal do ABC, Santo Andr\'e, S.P., Brazil, {ifialk@gmail.com}}
\and 
Maria Perel
\footnote{St. Petersburg State University, 7/9 Universitetskaya nab., St. Petersburg 199034, Russia,  {m.perel@spbu.ru}}
 }


\maketitle
\begin{abstract}
 An asymptotic approach for a Schroedinger type equation with non selfadjoint Hamiltonian of a special type in the case of two close degeneracy (turning) points is developed. Both real and complex degeneracy points are treated by a method of matched asymptotic expansions in the context of a unifying approach.
An asymptotic expansion near degeneracy point containing the parabolic cylinder functions is constructed and the transition matrix connecting the coefficients of adiabatic modes in front of and behind the degeneracy point is derived.

A simple non-technical recipe is also provided, which enables one to apply results to different physical 
problems without  performing  intermediate calculations.

{Keywords: degeneracy points, turning points, avoided crossing, non-Hermitian Hamiltonians}
\end{abstract}

\section{Introduction}\label{sec:WKB}

Formal asymptotic solutions of a Schroedinger type equation
\be
	  \H(x)  \Ps(x) = -i \h \frac{\partial \Ps(x)}{\partial x},
\label{Schr0}
\ee
where $ \H$ is a selfadjoint linear operator and $\h$ is a small parameter, $\h \to 0$, have been studied intensively from the beginning of the XXth century. Interest in such an  equation was initially invoked and greatly stimulated by  problems of quantum mechanics, where such an equation describes the adiabatic evolution of a quantum system the  Hamiltonian of which  is given by $ \H$ and $-x$ stands for the physical time. For  $\h$ sufficiently small, the exact solution is approximated by an adiabatic, or semiclassical, expansion (which we also call an adiabatic mode) having the  principal term
\be
	\Ps = \vph \,e^{\tfrac{i}\h\int^x\b(x)}  + O(\h).
	\label{sWKB}
\ee
Here $\b$  is an eigenvalue of the operator $ \H$, and $\vph$ is a properly chosen corresponding eigenfunction,
\be
	  \H (x) \vph(x) =\b(x) \vph(x).
	\label{sEVP}
\ee
 It is implied here that $\vph$ comprises the factor containing the Berry phase if it is nontrivial.
The asymptotics \Ref{sWKB} works well if there is a finite gap between the eigenvalue $\b$ and  the rest of the  spectrum of $ \H$ and if this gap does not depend on $\h$.
 A discussion of the asymptotics that approximate exact solutions with an error $O(\h)$, $O(\h^n)$ with an integer $n$ or $O(e^{-c/\h})$  with constant $c$ proportional to the gap between the eigenvalue $\b$ and the rest of the spectrum is given in papers \cite{born}, \cite{garrido} and \cite{Nenciu93}, respectively. All different results in this field are known under a generalized name of {\it adiabatic theorem.}

\begin{figure}
\centering
 \includegraphics[width=13cm]{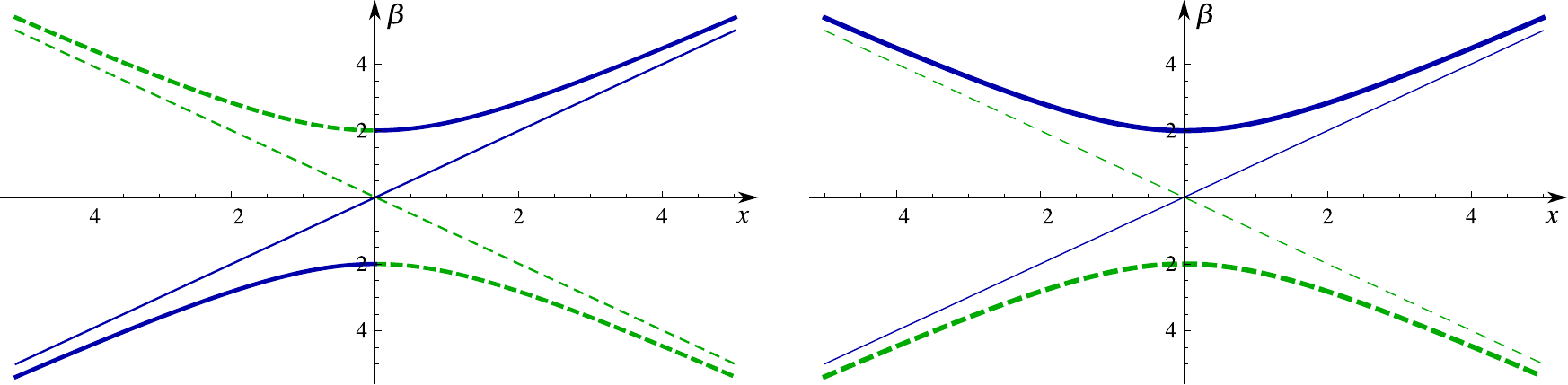}
\caption{
Avoided crossing: The nonperturbed eigenvalues (thin lines) are crossing at $x=0$, while the exact ones (thick lines) do not.  Dashed/solid lines (also blue/green on-line) distinguish different modes, left/right  pictures represent different numbering of exact eigenvalues, see Section \ref{sec:numbering}.  (In arbitrary units.)
\label{compl_TP}}
\end{figure}

It is well known that шт the vicinity of a crossing point  of at least two  eigenvalues $\b_j(x)$, $j=1,2$, say, at $x=0$ (thin lines  in Fig.~\ref{compl_TP}), the adiabatic approximation (\ref{sWKB}) is not applicable.
Neither is it applicable in the case of  an {\it avoided crossing} (see thick lines in Fig.~\ref{compl_TP})
  \be\label{av-cr}
 \frac{\b_2 - \b_1}{2} \simeq \sqrt{ (Q x)^2 + p^2 },
 \ee
which appears upon a small perturbation of the operator with  crossing  eigenvalues; see, for example, \cite{qm_landavshitz_59}.
The parameter $p$ in \Ref{av-cr} characterizes the smallest distance between the eigenvalues. If $p$ decreases rapidly enough with  ${\h}$, the adiabatic approximation \Ref{sWKB} does not work near $x=0$.

For the first time, the problem of constructing  asymptotic solutions in the presence of an avoided crossing  was encountered in quantum mechanics in the description of state transitions in a two level quantum system. The transitions amplitudes were found by Landau \cite{landau_32} and Zener \cite{zener_32} in 1932. The refined results were obtained by Stueckelberg \cite{stueckelberg_32} shortly afterwards.
Investigation of the phase of the transition coefficients were performed in \cite{bobashov}, \cite{bobashov2}. Among many other works on the adiabatic transitions in physical systems, some are to be outlined including the Dykhne's formula \cite{Dykhne62}, Hwang  and Pechukas, e.g. \cite{Hwang77}, and Zhu and Nakamura \cite{Zhu94}.
%
 However, mathematically rigorous results came only much later, and the first proof of the Landau-Zener formula  was given in \cite{LZ_hagedorn_91,LZ_joye_94}.  The Landau-Zener  formula was derived for a system of pseudo-differential equations in \cite{verdier_99}.

\begin{figure}
\centering
 \includegraphics[width=13cm]{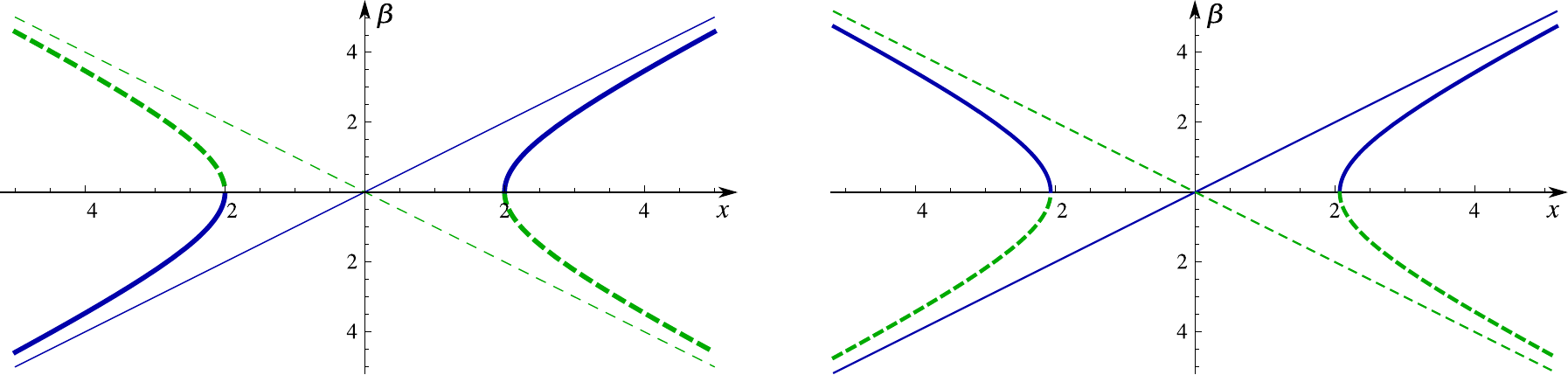}
\caption{
Real turning points: Non perturbed eigenvalues (thin lines) are crossing at one point, $x=0$, as before. However, the exact ones (thick lines) are  crossing at two points separated by  perturbation. In between the turning points, the eigenvalues are purely imaginary. Dashed/solid lines (also blue/green on-line) distinguish different modes, left/right  pictures represent different numbering of exact eigenvalues, see Section \ref{sec:numbering}. (In arbitrary units.)
\label{real_TP}}
\end{figure}

Based on \Ref{av-cr} it was reasonably stated in \cite{landau_32} that an avoided crossing is actually a crossing of the eigenvalues at two points in the complex plane $x=\pm ip/Q$.
At the same time, the crossing of  eigenvalues at a real point, which cannot be resolved by a small perturbation, is also widely known in the theory of the so-called WKB approximations \cite{wkb_wentzel_26,wkb_kramers_26,wkb_brillouin_26}. Primarily, it happens in the problems described in terms of a  second order differential equation, such as the stationary Schroedinger equation itself,
\be
	-\frac{\hbar^2}{2m}\psi''(x)+ (V(x)-E)\psi(x)=0.
\label{Schr-pure}
\ee
In the  WKB approximation,  $\b_{1,2}\equiv\pm \sqrt{E-V(x)}$ plays the role of the eigenvalues \Ref{sEVP}.  The solutions in this case are given by an expression similar to \Ref{sWKB}. In the 	classically allowed region, i.e., where $V<E$, the solution with a positive eigenvalue, $\b_1$, is interpreted as a running forward  along the $x$ axis, while the solution with a negative $\b_2$ is running backward. The points $\varkappa$, where the eigenvalues vanish, $E=V(\varkappa)$, and thus degenerate, $\b_1=\b_2=0$, are called \emph{turning points}. It is in their neighborhood  that the WKB approximation is not valid anymore, which resembles  the behavior of adiabatic solutions \Ref{sWKB} for equation (\ref{Schr0}). Near turning points, the eigenvalues depend on $x$ as $\b_j \simeq (-1)^j \sqrt{-V'(\varkappa)(x-\varkappa)}$, $j=1,2$, if $V'(\varkappa)\ne 0$.  Turning points with such eigenvalue behavior  are called the \emph{simple} ones. If $V=V(x)$ has  a single absolute extremum on interval of interest not coinciding with $\varkappa$, there are generally two simple turning points, see Fig.~\ref{real_TP}.

 The general solution of  \Ref{Schr-pure} in the presence of one simple turning point comprises two oscillating asymptotics on one side (classically allowed region) and an exponentially decreasing and increasing ones on the other (classically forbidden region). A particular solution, which has only exponentially decreasing asymptotics,
 may be interpreted as follows:  a wave incident to a such point is completely reflected, hence the name of the  turning  point.

However, for some  systems of ordinary differential equations (ODEs) and for more general cases, the interpretation of the point, where two eigenvalues degenerate, as a point of reflection is not valid. Therefore we prefer the terms ``the point of degeneration" or the \emph{degeneracy point}, using the other names only emphasizing the character of degeneracy. For uniformity of the nomenclature, we refer to an avoided crossing as two simple complex degeneracy points. 

The $x$-dependence of eigenvalues near the degeneracy point does not necessary contain the square root. It may be as $\b_{1,2} \sim \pm (x-\varkappa)^{k/2}$ for any integer $k$.
Every value of $k$ demands a special investigation. Below we deal with the perturbation of an operator with crossing eigenvalues, i.e., with one degeneracy point, near which  the eigenvalues are linear functions ($k=2$).  We show that after perturbation such a degeneracy point splits into a pair of simple ones.

The problems of constructing of asymptotic solutions in the presence of one degeneracy point or a pair of them for the ODEs of second order and their generalizations to systems of ODEs has been studied in many papers;  see the books and  reviews \cite{berry72}, \cite{Fedoryuk}, \cite{Olver}, \cite{Slavyanov}, \cite{turn_waso_85} and references therein. All the methods applied to investigation of degeneracy points can be divided into three groups:  the uniform methods, which work both in the vicinity of degeneracy points and away from the points, methods based on the Fourier representation of the unknown function, and methods based on local considerations in the vicinity of the degeneracy point with further matching of local solutions with adiabatic ones.

In the technique of `uniform approximations' the required solution is mapped into a solution of a simpler equation, which, however, has the same disposition of degeneracy points as the original one. It was pioneered in
 \cite{Buldyrev-Slavyanov}, \cite{Cherry}, \cite{un_ass_langer_37}.
 It was also applied to the equations in the general form in \cite{buslaev_grigis_01}. The methods based on the Fourier analysis are the Maslov method \cite{Kucherenko}, \cite{Maslov} and the microlocal analysis  \cite{verdier_99}. In the next section, we give a more detailed exposition of the  method of matched asymptotic expansions (or the boundary layer method), which we apply in this paper and which is based on local considerations.

An abstract general approach to degeneracy points for equation \Ref{Schr0}, where  $\H(x)$ is an operator in the Banach space has been suggested in  \cite{buslaev_grigis_01}, \cite{Grinina}, see also references therein. A spectral model equation of the same form \Ref{Schr0} is introduced in these works, where  $\H(x)$ is substituted with a matrix $2\times2$, with eigenvalues coinciding with eigenvalues of the original problem. Despite of the  generality of the approach, only three types of spectral models were considered: the \emph{parabolic} model with one simple turning point, the \emph{hyperbolic} model with a pair of turning points in between of which the eigenvalues are complex and real outside, and the \emph{elliptic} model with complexity of the eigenvalues inverted in regard to the pair of turning points.
The models with complex degeneracy points have  not been considered.

The treatment of the problem in \cite{buslaev_grigis_01} has the main emphasis on the mathematical rigor and validity of the approach. Hence, it does not contain a recipe of solving specific physical problems given by some particular ODE systems.

On the other hand, there is a lot of physical works treating different physical phenomena caused by the presence of degeneracy points. These physical works apart from the problems for Schroedinger equation comprise, for example, the problems of radiowaves propagation \cite{ion_bitchutskaja_79},  \cite{Budden72}, \cite{ion_budden_75}, \cite{ion_perel_90},  acoustical waveguides \cite{wg_molotkov_78,wg_molotkov_74},   elasic waveguides \cite{perel_kaplunov_05},  \cite{Pagneux-Maurel}, waves in the Timoshenko beam \cite{timo_perel_00},  electrons in crystal \cite{Buslaev-Dmitrieva_87},   liquid crystals \cite{Aksenova}, graphene \cite{koen}, \cite{zalip},  creeping waves on the smooth boundary of the convex body for particular boundary conditions  \cite{And-Zai-Per} to mention just a few. Despite profound similarities of all the problems with degeneracy points, each physical problem from above was solved with account of its specific assumptions and without applying any of the theory developed in the abstract form in the earlier works.

\vskip5mm

It is the intention of the present paper to fill the existing gap between abstract considerations of \cite{buslaev_grigis_01}, \cite{Grinina},
 and the needs of the physical community dealing with  particular problems. We build a mathematical approach, which permits us to consider in the general form and on the same footing pairs of both real and complex degeneracy points generated by a small perturbation of an operator with two eigenvalues exactly degenerate at one point.

Our approach is based on some
  spectral properties of a selfadjoint linear operator  pencil, to which most of the relevant  problems can be reduced. It originates from the consideration of the  Schroedinger equation with a not selfadjoint Hamiltonian of a special type. The classification of different problems is much easier in this formulation. Apart of the original eigenvalue behavior, the classification requires  an additional condition, 
which turns out to be a linear independence of the corresponding eigenvectors.

Along with deriving an asymptotic solution for the problem at hand, we also aim to deliver a (relatively) nontechnical recipe applicable to a wide range of physical problems. The objective is to give the resulting formulas in such a form that the understanding of intermediate procedures and details of their derivation are not necessary for their application.

Away from the degeneracy point, such an approach for waves propagation problems has been  first  applied in \cite{felsen_marcuvitz_72}. To the best of our knowledge, it was in  \cite{ion_perel_90} and \cite{timo_perel_00}, where such a statement has been pioneered for considering particular physical problems in the presence of degeneracy points.


\section{Statement of the problem and outline of the paper}\label{sec:stat-pr}

To obtain the necessary level of rigor and generality of our considerations but to maintain at the same time the transparency of derivation and guarantee the applicability to a wide range of physical problems, we elaborate a method of the not selfadjoint Schroedinger equation of a special type.

The suggestive observation is that \Ref{Schr-pure} can be written in the form of \Ref{Schr0} (see App. \ref{app:examples}) but with a non selfadjoint $ \H$, which, however, can be factorized as
\be\label{factor}
	 \H = \G^{-1} \hK,
\ee
where both $\G$ and $\hK$ are selfadjoint. Thus, we focus our considerations on the following linear operator equation
\be
	\hK \Ps(x) = -i\h \G\frac{\partial \Ps(x)}{\partial x},
	\label{newSch}
\ee
where both $\hK$  and $\G$ are selfadjoint operators in an appropriate Hilbert space, to which $\Ps$ belongs, $\G$ and $\G^{-1}$ are bounded. The operator $\hK$ may  depend on other coordinates as well as $x$, but it should not have derivatives in its respect. We assume that $\G$ does not depend on $x$. This condition may be removed but we include it for simplicity, because in the cases known to the authors it is always satisfied.  Equation \Ref{newSch} is a direct generalization of \Ref{Schr0} and for non trivial $\G$ it is known in mathematical literature as of Sobolev type, see for example, \cite{sviridyuk_94}.

Further analyses shows that most of the physical problems with degeneracy points can be written in form \Ref{newSch}. In particular the stationary Schroedinger equation \Ref{Schr-pure},  Maxwell equations \cite{ion_perel_90} (with a non-invertible $\G$), the Timoshenko beam equations \cite{timo_perel_00}, Dirac equation for fermions scattering in graphene, and many others, see also examples of Appendix \ref{app:examples}. Actually, any system of linear differential equations can be written in a similar form and, if it describes a physical process with the conservation of energy,  the operators $\hK$ and  $\G$ are always selfadjoint according to our observations.
Equation \Ref{newSch} can also be treated in the framework of $\mathcal{PT}$-simmetric quantum mechanics with $\G$ playing the role of $\mathcal P$-simmerty operator, see \cite{Bender2007} and reference therein.

One of the important features of \Ref{newSch} is that it possesses a conservation law similar to the total probability conservation for the Schrodinger equation. For any two  exact solutions of (\ref{newSch}) $\Ps_1$ and $\Ps_2$ it holds%
\footnote{Henceforth by $(\cdot,\cdot)$ we understand the standard scalar product of the corresponding Hilbert space.}
\be \label{conser}
 (\Ps_1,\G \Ps_2) = const.
\ee
Indeed, differentiating the left-hand side of (\ref{conser}) we deduce
\be
\frac{\partial (\Ps_1,  {\G}\Ps_2)}{\partial x}
	=(\Ps_1, {{\G}}\frac{\partial \Ps_2}{\partial x}) + ({\G} \frac{\partial \Ps_1}{\partial x}, \Ps_2) =
 -i\h^{-1}(\Ps_1,\hK \Ps_2) + i\h^{-1}(\hK \Ps_1, \Ps_2) = 0,
\ee
where we used that operators $\hK$ and $\G$ are self-adjoint.
For $\G=\I$, identity operator, the constant in \Ref{conser} is simply a normalization constant and \Ref{conser} gives the familiar total probability conservation law of quantum mechanics. In the problems of waves propagation from Appendix \ref{app:examples} we have $\G\ne\I$ and the quantity (\ref{conser}) has a meaning of time-averaged flux density in the $x$-direction, while for the Dirac equation it is a time-averaged electron current density in the same direction. In this paper we will refer to \Ref{conser} simply as \emph{flux}. Note, that the constant in \Ref{conser} is not positive definite, see more below.

\vskip5mm

To take into account the presence of (small) perturbation in many relevant physical problems, we consider a particular form of $\hK$,
\be \label{perturbed-Ham}
	\hK = \K(x) + \delta \B(x), \quad \delta = \sh,
\ee
thus giving the subject of our interest in the form
\be
	\(\K+\sqrt\h \B\)\Ps= -i\h \G\frac{\partial \Ps(x)}{\partial x}.
	\label{mainEq0}
\ee
One could consider a more general two-parametric case, with $\delta$ independent of $\h$, but we choose the most important relation between them to avoid complications that do not affect the main result. The importance of the relation $\delta=\sh$ was known since the early works of Landau \cite{landau_32} and Zener \cite{zener_32}. The two-parametric consideration of the case, where $\G$ is the identity matrix, was given in \cite{verdier_99} and confirmed the importance of this relation.

Further, we assume that the operators $\G$ and its inverse, $\B(x)$, and $\K(x) - \K(0)$ are all bounded.
We assume also that $\K(x) - \K(0)$ and $\B$  may be expanded in a series in powers of $x$ near $x=0$.

The principal term of the adiabatic solution for \Ref{mainEq0} is very similar to \Ref{sWKB} and is constructed by means of an eigenvalue $\b$ and eigenfunction $\vph$ of the spectral problem given by
\be
	 \K(x) \vph(x) =\b(x) \G \vph(x).
	\label{newEVP}
\ee
Eigenvalues of \Ref{newEVP} can be both real and complex, while eigenfunctions $\vph$ are $\G$-orthogonal ( see   Appendix \ref{app:eig-pr} for details).

We make two crucial assumptions on the behavior of $\b$ and $\vph$.
\begin{enumerate}
\item{
We limit ourselves to consideration of the case where two real eigenvalues $\b_1$ and $\b_2$ of \Ref{newEVP} have a point of simple crossing at $x=0$, i.e.,
\be
	 \b_2(x) -  \b_1(x) \seq_{x\to0} 2 Q x,\qquad
	Q>0.
	\label{mb-cross}
\ee
$Q$ does not depend on $\h$, $Q \sim 1$, and both  $\b_1$ and $\b_2$ are separated from the rest of the spectrum of  \Ref{newEVP} (if any) with a gap independent on $\h$. This condition also fixes the numbering of the unperturbed modes, for a detailed discussion of this issue see Section \ref{sec:numbering}.
}

\item{
The corresponding eigenfunctions $\vph_j(0)\equiv \,\lim_{x\to 0} \vph_j(x)$, $\, j=1,2$,  are linear independent at $x=0$. }
\end{enumerate}
In fact, we assume that the eigenvalues $\b_j(x)$ and eigenfunctions $\vph_j(x)$ can be  found as series in powers of $x$.

However, the eigenvalues and the eigenfunctions $\b_j$, $\vph_j$, $j=1,2$, are holomorphic functions of $x$ and the series for them converges   if $\K(x)-\K(0)$ and $\B(x)$ are holomorphic families of operators, and the eigenvalues $\b_j$, $j=1,2$, are separated from the rest of the spectrum with a finite gap \cite{Kato}.

Our aim is to find formal asymptotic expansion of solutions \Ref{mainEq0} as $\sh \to 0$ both away from the degeneracy point and in its neighborhood. We solve also the connection problem.  Its formulation is given in Section \ref{sec:numbering}.

The case of linear  dependent $\vph_1(0)$ and $\vph_2(0)$ will be considered in the next paper.  We note that this condition does not hold for equation (\ref{Schr-pure}), which  was our original motivation for this paper.  However it is fulfilled for the important physical problem of the Dirac fermions \cite{koen}, among other examples, see Appendix \ref{app:examples}.

As will be shown in the next Section, our formulation of the problem incorporates both the case of an avoided crossing and  classical turning points. The two regimes are distinguished by the relative sign of the normalization of modes,
\be
	{\rm sign\ }[(\vph_1,\G\vph_1)(\vph_2,\G\vph_2)] = \pm1,
\label{conser2}
\ee
which is an invariant for a given physical problem. 

Another spectral problem naturally arising in connection with \Ref{mainEq0} is the problem  for the whole operator $\hK$,
\be\label{newEVP1}
	(\K+\sh \B )\vvph = \hb \G \vvph.
\ee
The eigenvalues $\hb$ and eigenfunctions $\vvph$ will be found in Section \ref{sec:EVP-close}
 as perturbations of the eigenvalues $\b$ and eigenfunctions $\vph$ of \Ref{newEVP}.

\vskip5mm

In this paper, we follow the method of matched asymptotic expansions \cite{bender_book,turn_waso_85},  also called the boundary layer method  \cite{boundl_babich_79}.  It consists of the construction, in the vicinity of a turning point, where the adiabatic (also called the {\it outer}) expansion is not applicable, of an \emph{inner expansion}, which is given in terms of a series expansion in particular powers of a small parameter, $\h^\a$, $\a<1$. The validity zones of two types of solutions intersect, and they can be matched there to obtain the transition matrix. In the field of nonadiabatic transitions, this method has been used for the first time by Hagedorn in \cite{LZ_hagedorn_91} for his proof of the Landau--Zener formula. Methodologically, our paper follows the asymptotic considerations of our previous works \cite{And-Zai-Per,ion_perel_90,timo_perel_00,perel_kaplunov_05}.


The paper is organized in the following way. The Introduction and the Statement of the problem is followed by   Section \ref{sec:EVP}, which  contains auxiliary, though very important considerations of eigenvalues and eigenfunctions of $\hK$ near a degeneracy point, i.e., the perturbed eigenvalue problem. The formulas obtained there enable us to introduce all the parameters, which determine our final results.  Namely, in this section  we derive that the eigenvalues have complex degeneracy
 points (avoided crossing case), Fig.~1, or the real ones, Fig.~2, (classical turning points)  depending on the relative sign of normalization of two degenerating modes, as stated above.

In Sections \ref{sec:outer}--\ref{sec:T}, we realize the program of  matched asymptotic expansions, starting with the construction of the adiabatic (outer) expansion in Section \ref{sec:outer}. It maintains its asymptotic character outside a neighborhood of the degeneracy point, whose size we derive in Section \ref{sec:wkb-valid}. The analysis of the outer expansion on the boundary of its applicability zone enables us to introduce a stretched variable.

Upon constructing the adiabatic solutions we define in Section \ref{sec:numbering}  the transition problem between them and discus their numbering at different sides of the degeneracy point. After that, in Section \ref{sec:inner} we rewrite the equation \Ref{mainEq0} in terms of the stretched variable and construct an inner expansion, which works near the degeneracy point. An inner expansion contains a special function; in our case it is the parabolic cylinder function.  The zones of validity of the outer and the inner expansions intersect as expected. We find an asymptotics of the inner expansion for a large stretched variable in Section \ref{sec:inner-matching}, and rearrange an adiabatic expansion in terms of the stretched variable in Section \ref{adia_rearr}.

Finally, we match these expansions in the intersection zone and obtain the transition matrix in Section \ref{sec:T}.
It naturally depends on the definition of the adiabatic modes. We introduce canonical adiabatic modes in Section \ref{sec:can-modes}, which provide the transition matrix in the simplest form and also give a general transition matrix for an arbitrary choice of the adiabatic modes in Section \ref{sec:arb-modes}.
It is followed by Section \ref{sec interp} where we give a physical interpretation of the obtain transition matrix in term of reflection and transition coefficients.

In Conclusions, section  \ref{sec:concl} of the paper, we formulate shortly our main results and give the recipe for those,  who do not want to delve in the process of obtaining  asymptotic formulas and want to jump directly to a result applicable to any particular problem, which can be reduced to an equation of the form  \Ref{mainEq0}.

The paper is concluded with four Appendices.  In the first one, Appendix \ref{app:eig-pr}, we discuss the basic features of the eigenproblem \Ref{newEVP} of selfadjoint linear operator pencils.  Some properties of the outer expansion in the limit of lifting degeneracy are outlined in Appendix \ref{app:outer-lift}. The general properties of the transition matrix following from the conservation law \Ref{conser} are discussed in Appendix \ref{app:T_prop}.  Appendix \ref{app:examples} contains some examples of problems, which are reducible to \Ref{mainEq0}, and a particular case of application of our method to electrons scattering in  graphene.


\section{Perturbation method for the spectral problem}\label{sec:EVP}

Since many of the particular problems of waves propagation or states transition are given in terms of the behavior of the spectrum, we start by investigating the spectral problem  \Ref{newEVP1}.
To this end we adapt the perturbation method developed by Schroedinger and Rellich (see the history of the problem and references in \cite{Kato}) to deal with the operator pencils rather then operators themselves, i.e., with eigenproblems containing $\G$ on the right-hand side.

 We construct an approximate solution to the eigenvalue problem both away from the degeneracy point and in its vicinity. Applying the perturbation method to find the eigenvalues, we also will be able
to introduce physically relevant parameters  governing our results, and
to clarify the conditions, which cause eigenvalues behavior either as shown in Figs. \ref{compl_TP} or \ref{real_TP}.

We use in this Section properties of eigenvalues and eigenfunctions of the  unperturbed spectral problem \Ref{newEVP} derived in Appendix \ref{app:eig-pr}.

\subsection{Perturbed eigenproblem away from degeneracy point}\label{sec:EVP-away}
Away from degeneracy point, we may search for the eigenvalues and eigenfunctions of \Ref{newEVP1}
$$
	(\K+\sh \B )\cvph = \vb \G \cvph
$$
as a formal asymptotic series in powers of $\sh$ and as functions of the  original variable $x$,
\be
	\vb(x,\h)={\vbn0}(x)+\sqrt\h{\vbn1}(x)+\ldots,\qquad
	\cvph(x,\h)=\cvphn0(x)+\sqrt\h \cvphn1(x)+\ldots
	\label{anz0}
\ee
This procedure is very well studied for $\G=\I$, and will mostly be used as the reference in the rest of the paper, so we just indicate the main steps postponing detailed treatment to further sections.

Upon inserting these series into \Ref{slVarEVP} and equating terms with equal powers of $\sh$, we get an infinite sequence of equations
\begin{eqnarray}
	(\K-\vbn0\G)\cvphn0&=&
		0\label{0-th-B}\\
	(\K-\vbn0\G)\cvphn1&=&
		 \( \vbn1\G-\B\)\cvphn0
		\label{1-th-B}\\
	(\K-\vbn0\G)\cvphn{2}&=&
		 \( \vbn1\G-\B\)\cvphn1 + \vbn2\G\cvphn0
		\label{2-nd-B}\\
&\ldots&\nonumber
\end{eqnarray}
Equation \Ref{0-th-B} is the original spectral problem \Ref{newEVP}. Thus, we choose the principal approximation as
\be
	\vb_{j}^{(0)}=\b_j, \qquad
	{\cvph}_j^{(0)} = \vph_j
	\label{EF}
\ee
where $j$ is the number of the corresponding solution of the eigenproblem, for the sake of definiteness we choose to construct here the first mode, $j=1$.

The solution of the above system of equations differs from the standard case
only in the definition of the unperturbed eigenvalues and eigenfunctions $\b$, $\vph$, and the presence of $\G$. Thus, in the first approximation we can write
\be
\cvph_1=\vph_1
	+\sqrt\h\[\frac{ \Bc_{ 21}}{( \b_1- \b_2)   N_2} \vph_2
		+ \check\vph_{1\perp}^{(1)}\]
	+\ldots
	\label{vvph}
\ee
\be
\vb_1=\b_1+\sqrt\h \frac{ \Bc_{ 11}}{  N_1}
 +\h\[\frac{  \Bc_{ 21}  \Bc_{ 12}}{ ( \b_1- \b_2)   N_1  N_2 }
	+\frac{ (\vph_1,\B \check\vph_{1\perp}^{(1)})}{  N_1} \]
	+\ldots,
	\label{hb}
\ee
We separated explicitly the contribution of the second mode $\vph_2$ since we assume that namely the pair of $\b_1$ and $\b_2$ degenerate at $x=0$.
 By $\vph_{1\perp}^{(1)}$ we denoted the contribution of the rest of the spectrum (if any) to the first approximation of $\cvph$, $\G$-orthogonal to $ \vph_1$ and $ \vph_2$.
  We also supplied the expansion \Ref{anz0} with condition
\be
	({\cvph}^{(0)},\G{\cvph}_j^{(n)})=0.
	\label{add-cond}
\ee
In \Ref{vvph}, \Ref{hb}, we also introduced the following notation for the  eigenfunctions normalization
\be
	N_i=(\vph_{i},\G\vph_i), \quad i=1,2.
	\label{N-def}
\ee
It can always be chosen constant.  The matrix elements for any operator, say, ${\bm{\mathcal A}}$, are defined as
\be
	{\cal A}_{ij}(x)\equiv \(\vph_i(x),{\bm{\mathcal A}}(x)\vph_j(x)\),\quad j,k=1,2.
	\label{matrix-ele}
\ee
The scalar product here is that inherent from the Hilbert space in which $\K$ acts.

\subsection{Perturbed eigenproblem in the vicinity of the degeneracy point}\label{sec:EVP-close}

As  is evident from \Ref{vvph}, \Ref{hb}, the expansions are not valid whenever $\b_j$, $j=1,2$, degenate.  Similarly  to the general theory \cite{boundl_babich_79,bender_book,turn_waso_85} for $\G=\I$, and other particular problems, e. g.,  \cite{timo_perel_00,perel_kaplunov_05},  in the vicinity of a degeneracy point the behavior of the eigenvalues/eigenfunctions  (and of the approximate solution of the equation \Ref{mainEq0}) should be described in terms of a \emph{slow variable}. For our particular case with perturbation of order $\sh$ and linear intersection of the eigenvalues, it is given by
\be
	\tau=x/\sh.
	\label{slVar0}
\ee
Arguments for choosing such a slow variable are given in Section \ref{sec:wkb-valid}.


We substitute   (\ref{slVar0}) in $\K$ and $\B$  and expand them in the formal series as follows
\be\label{expan-H-B}
 \begin{aligned}
	\K(\sh \tau)&= \Kn0    + \sh\tau \Kn1   + \h \tau^2 \Kn2+\ldots, &\\
	 \B(\sh \tau)&= \Bn0     + \sh\tau \Bn1   + \h \tau^2 \Bn2 +\ldots,&
 \end{aligned}
\ee
where
$$
	\K^{(n)}=\left.\frac1{n!} \frac{d^n \K}{d x^n}\right| _{x=0}
	, \quad \B^{(n)}=\left.\frac1{n!} \frac{d^n \B}{d x^n}\right | _{x=0}.
$$
Inserting these expansions into (\ref{newEVP1}) we obtain
\be
	\(\Kn0+\sh (\tau \Kn1 +  \Bn0) + \h (\tau^2 \Kn2 +  \Bn1\tau)+\ldots \) \vvph = \hb \G \vvph.
	\label{slVarEVP}
\ee
We search for the eigenvalues and eigenfunctions in the form similar to \Ref{anz0}
\be\label{anzats-pert-eig}
	\hb(\tau,\sh)={\hbn0}(\tau)+\sqrt\h{\hbn1}(\tau)+\ldots,\quad
	\vvph(\tau,\sh)=\vvphn0(\tau)+\sqrt\h \vvphn1(\tau)+\ldots
\ee
Note that in distinction to the expansions of $\vb$, $\cvph$, the approximations $\hbn{n} $, $\vvphn{n} $  are now functions of $\tau$,  not $x$. We distinguish them by using a hat accent.

Upon inserting these series into \Ref{slVarEVP} and equating terms with equal powers of $\sh$, we get an infinite sequence of equations
\begin{eqnarray}
	(\Kn0-\hbn0\G)\vvphn0&=&
		0\label{0-th-tau}\\
	(\Kn0-\hbn0\G)\vvphn1&=&
		 \( \hbn1\G-\tau\Kn1-\Bn0\)\vvphn0
		\label{1-th-tau}\\
	(\Kn0-\hbn0\G)\vvphn{2}&=&
		 \( \hbn1\G-\tau\Kn1-\Bn0\)\vvphn1\nonumber\\
		  &&+ \( \hbn2\G-\tau^2\Kn2-\tau\Bn1\)\vvphn0
		\label{2-nd-tau}\\
&\ldots&\nonumber
\end{eqnarray}
Equation \Ref{0-th-tau} is the original spectral problem \Ref{newEVP} taken at the degeneracy point $x=0$. Thus we have
\be
	\hb_{j}^{(0)} = \b_0,
\ee
where
\be
\b_0 \equiv \b_1(0)=\b_2(0), \label{b-0-def}
\ee
and  we introduced the subscript, $j=1,2$, to distinguish the two modes in what follows.
For the eigenvalue we have
\be
{\vvph}_j^{(0)} =  \a_{j1}^{(0)}(\tau) \vph_1(0) + \a_{j2}^{(0)}(\tau) \vph_2(0), \quad j=1,2,	\label{EF}
\ee
where $ \vph_{j}(0) \equiv \lim_{x \to 0}  \vph_{j}(x)$, $j=1,2$  are linear independent eigenfunctions of the original problem. We determine them by the limiting transition to make them uniquely defined.
The eigenfunctions of  both degenerating modes may be written in the form \Ref{EF}.

The coefficients $\a_{jk}^{(0)},$ $j,k=1,2$ are unknown at this step. They can be found from the condition of the solvability of the  equation \Ref{1-th-tau}, which is the condition of orthogonality of the right-hand side of \Ref{1-th-tau} with $ \vph_k(0)$, $k=1,2$.
It gives for both  $j=1$ and $j=2$ the same system
\be
\begin{aligned}
	(\hb_{j}^{(1)} N_1^{(0)} - \tau \Kcn1_{11} - \Bcn0_{11})\, \a_{j1}^{(0)}
		+ (-\tau \Kcn1_{12} - \Bcn0_{12}) \, \a_{j2}^{(0)}&=0,&\\
	 (-\tau \Kcn1_{21}- \Bcn0_{21}) \, \a_{j1}^{(0)}
	+(\hb_{j}^{(1)} N_2^{(0)} - \tau \Kcn1_{22} - \Bcn0_{22}) \, \a_{j2}^{(0)}&=0;&
\end{aligned}
\label{eq-eig-pert}
\ee
here, $N_j^{(0)}=N_j(0)$, $j=1,2$. Similarly, the matrix elements with a superscript $(n)$, $n=0,1,$ are also taken at the degeneracy point $x=0$,
\be
\Kcn1_{jk} = (  \vph_j(0),  \K^{(1)}(0)   \vph_k(0)), \qquad
	\Bcn0_{jk} =  \Bc_{\brr jk}(0), \quad j,k=1,2.
\ee
Note that, as we show in the Appendix \ref{app:eig-pr}, $\Kcn1_{12}=\Kcn1_{21}=0$  always holds. The diagonal matrix elements $\Kcn1_{jj}$, $j=1,2,\ldots$, are proportional to the derivatives of the unperturbed eigenvalues, see \Ref{db-dx-h}.

The condition of solvability of the system \Ref{eq-eig-pert} with respect to $\a_{j1}^{(0)}$,  $\a_{j2}^{(0)}$  is the  nullification of the determinant. It gives a quadratic equation in  $\hb^{(1)}_j$. Under an appropriate choice of the notation, its solution can always be written as%
\be\label{eig-1app}
	\hb^{(1)}_{1,2} = \hb_{av}^{(1)} \pm \sqrt {(\tau + b)^2 Q^2
			+   {p^2}{\rm sgn }(N_1^{(0)} N_2^{(0)}) }
\ee
The square root is assumed to be positive if it is real. We do not discuss the complex case since we will not treat it.

In \Ref{eig-1app}, we used the following notation.
Half  the difference of the derivatives of the unperturbed eigenvalues at $x=0$ is denoted as
\be
	\label{not-Q}
	Q = \frac{1}{2}	\(\frac{\Kcn1_{22}}{  N_2^{(0)}}  - \frac{\Kcn1_{11}}{  N_1^{(0)}} \), \qquad  Q>0,
\ee
and the degeneracy point displacement owing to the perturbation equal for both modes reads
\be \label{not-b}
  b = \frac{1}{2Q} \( \frac{\Bcn0_{\brr 22}}{  N_2^{(0)}} - \frac{\Bcn0_{\brr 11}}{  N_1^{(0)}}\).
\ee
The parameter $p$ characterizes the degree of separation of the eigenvalues for $\sgn (N_1^{(0)} N_2^{(0)}) \equiv \sgn (N_1N_2) =1 $
\footnote{As shown in Appendix \ref{app:eig-pr}, the norms of eigenfunctions $\vph_j$, $j=1,2$, cannot vanish, once eigenfunctions are smooth and linearly independent on the whole interval. Then the sign of the norm of $\vph_j$, ${\rm sgn}( N_j )$, is a constant even for $N_j=N_j(x)$, and so is the product, ${\rm sgn}(N_1^{(0)} N_2^{(0)}) = {\rm sgn}(N_1 N_2) $. In what follows we always use the later expression.}, or the width of the classically forbidden zone if $N_1 N_2 <0$.
It reads
\be
p^2 = \frac{\Bcn0_{12} \Bcn0_{21}}{|N_1^{(0)} N_2^{(0)}|}
	= \frac{|\Bcn0_{12}|^2}{|N_1^{(0)} N_2^{(0)}|}.
	\label{not-p}
\ee
The parameter $\nu$ is a dimensionless combination of the above mentioned physical parameters, which will govern our final result
\be
	\nu = i \frac{\Bcn0_{12} \Bcn0_{21}}{N_1^{(0)} N_2^{(0)}} \frac{1}{(\,\b_2'(0)-\b_1'(0)\,)}
		= \frac{i p^2\, {\sgn}(N_1 N_2)}{2 Q}.
	\label{nu_gen}
\ee
In what follows we also need $\sqrt{\nu}$. We define the branch of the square root in such a way that
\be\label{nu_gen_s}
\sqrt{\nu} = e^{i \frac{\pi}{4} {\rm sgn}(N_1 N_2 )}\sqrt{|\nu|}.
\ee
Finally, the average of degenerating eigenvalues in the first-order approximation is as follows:
\be\label{b-av}
\hb_{av}^{(1)}
	= \frac{1}{2} \(\frac{\Bcn0_{11}}{  N_1^{(0)}} + \frac{\Bcn0_{\brr 22}}{  N_2^{(0)}}\)
		+ \frac{\t}{2} \( \frac{\Kcn1_{11}}{  N_1^{(0)}} + \frac{\Kcn1_{22}}{  N_2^{(0)}} \).
\ee
Actually, by a simple transformation of the equation \Ref{newSch},
$ \Ps\to\Ps e^{\frac{i}{2 \h}\int^x {(\hb_1+\hb_2)}\, dx}$,
 we can always transform the operator $\hK$ to such a form that $\hb_1(x)=-\hb_2(x)$, and consequently $\b_0=\hb_{av}^{(1)}=0$. But we will keep our considerations in the general form.

We may give now formulas for principal approximations of eigenvalues:
\be\label{eig-root}
 \hb_{j} = \b_0 + \sh \( \,\hb_{av}^{(1)}\, + \,
   (-1)^j \,  \sqrt {(\tau + b)^2 Q^2
		+{p^2}\,\sgn (N_1N_2)  }\, \) + O( \h)
 \ee
$$
= \b_0 + \sh \(\hb_{av}^{(1)} \, + \,
   (-1)^j \,\sqrt {(\tau + b)^2 Q^2
		- 2 iQ \nu }\, \)+ O( \h),\quad
		j=1,2.
$$

At this stage we are ready to deduce that for $\sgn(N_1 N_2 )=1$ we have the avoided crossing case, see Fig. \ref{compl_TP}, with two complex degeneracy points, $\tau_\pm=\varkappa_\pm/\sh$,  while for $\sgn (N_1 N_2 ) =-1$  we have two real ones, as in Fig. \ref{real_TP},
\be
\tau_{\pm} = \varkappa_{\pm}/\sh
	=\left\{
	\begin{array}{ll}
		-b \pm i p/Q,\ &N_1 N_2 >0\\
		-b \pm p/Q, &N_1 N_2 <0
		\end{array}
		\right.
		.
\label{deg-point}
\ee
In both cases the degeneracy points are the simples ones.

The eigenfunction approximation ${\vvph}_j^{(0)}$ \Ref{EF} for $j=1,2$ can be easily found now by solving \Ref{eq-eig-pert} for $\a_j^{(0)}$, $j=1,2$, which can be written in the notation \Ref{not-Q}, \Ref{not-b}, \Ref{b-av},  either as
\be\label{al-s}
	 \a_{j1}^{(0)}(\t) = \frac{\Bcn0_{12}}{N_1^{(0)} },\qquad
		 \a_{j2}^{(0)}(\t) = \hb_{j}^{(1)}-\hb_{av} + Q(\tau + b),
\ee
or
\be
 	 \a_{j1}^{(0)}(\t) = \hb_{j}^{(1)} -\hb_{av} - Q(\tau + b),\qquad
		 \a_{j2}^{(0)}(\t) =  \frac{\Bcn0_{21}}{N_2^{(0)} }.		
\ee
The solvability of \Ref{eq-eig-pert} guarantees that these two expressions differ only in the overall normalization of the principal approximation \Ref{EF} of the eigenfunction $\vvph_j$.

 Let us find an asymptotics of eigenvalues and eigenfunctions \Ref{eig-root} and \Ref{al-s} as $|\t| \to \infty$.
  The eigenvalues have the following asymptotics:
 \be
 \label{lar-t-as}
 \hb_j \seq
\b_0
	+ \sh \(\hb_{av}^{(1)}
		+ (-1)^j  \( Q\,|\t + b| -\frac{i \nu}{|\t| } \) + o(\t^{-1})  \) + O(\h).
 \ee
 We neglect $b$ in comparison with $\t$ in the denominator of the second term and use \Ref{nu_gen} for $\nu$.
 Taking into account  formulas  \Ref{not-Q}, \Ref{not-b}, \Ref{not-p}, \Ref{b-av}, from \Ref{lar-t-as} for $(\t+b)>0$ we obtain
 \be\label{b-non-pert}
 \hb_j \seq_{\t\to+\infty}  \b_0 + \frac{\Kcn1_{jj}}{  N_j^{(0)}} \sh \t
 	+ \sh \frac{\Bcn0_{jj}}{  N_j^{(0)}}
 		- \sh
 		\frac{i\nu}{\tau}
 		+ \ldots
 \ee
and for $(\tau+b)<0$ we get
\be
\begin{aligned}
 &\hb_1 \seq_{\t\to-\infty}  \b_0 + \frac{\Kcn1_{22}}{  N_2^{(0)}} \sh \t
 	+ \sh \frac{\Bcn0_{22}}{  N_2^{(0)}}
 		- \sh
 		\frac{i\nu}{\tau}
 		+ \ldots,\quad\\
 &\hb_2 \seq_{\t\to-\infty}  \b_0 + \frac{\Kcn1_{11}}{  N_1^{(0)}} \sh \t
 	+ \sh \frac{\Bcn0_{11}}{  N_1^{(0)}}
 		- \sh
 		\frac{i\nu}{\tau}
 		+ \ldots.
 \end{aligned}
 \ee
Formula \Ref{b-non-pert}
helps  to see that the perturbed eigenvalues $\hb_j$, $j=1,2$, are numbered in \Ref{eig-root} to approach the values of the unperturbed ones (see \Ref{mb-cross}, \Ref{hb}) to the right of $x=0$
\be\label{hb-b-t+}
	\hb_j  \mathop\to_{\tau\to+\infty} \b_j,\qquad
		j=1,2.
\ee
While to the left, the numbering of the perturbed eigenvalues is inverted
\be\label{hb-b-t-}
	\hb_1  \mathop\to_{\tau\to-\infty}  \b_2,\qquad \hb_2  \mathop\to_{\tau\to-\infty} \b_1
\ee
We note here that there are two possible ways on numbering of $\hb$ on different sides of the turning point provided that the numbering of $\b$ is fixed by \Ref{mb-cross}. The one adopted in \Ref{eig-root}, and leading to \Ref{hb-b-t+}, \Ref{hb-b-t-}, is most convenient for $N_1 N_2>0$, since it preserves the continuity of $\hb$ across the degeneracy points area, see thick lines on Fig. 1, right. However it does not respect the flux sign of modes in the case of $N_1  N_2 <0$, since ${\sgn} N_i$, $i=1,2$ is constant across the degeneracy point, see Appendix \ref{app:eig-pr}.

On the other hand, if one chooses the numbering of $\hb$ in such a way that $\hb_j\sim\b_j$ at both sides of the degeneracy point, as on Figs. 1 and 2, left, i.e.
\be
	\hb_j\mathop\to_{x\to\pm \infty}\b_j,\quad j=1,2
	\label{other-num}
\ee
the sign of the flux of every mode $\hPs$ will be conserved for $N_1  N_2 <0$. However, in this case the perturbed eigenvalues $\hb$ become non-smooth functions of $\tau$ at  $\tau=-b$ point for  $N_1  N_2 >0$. See further discussion of this issue in Section \ref{sec:numbering}.


\section{Adiabatic (outer) expansion}\label{sec:outer}

To proceed with resolving the connection problem via the method of matched asymptotic expansions, in this section we construct adiabatic, or outer, expansions away from the degeneracy points of the original equation
\be
	\hK\Ps\equiv (\K + \delta \B)\Ps=-i\h\partial_x\G\Ps;
	\label{mainEq}
\ee
here $\partial_x\equiv \frac{\partial}{\partial x}$. As discussed in the Introduction, we limit ourselves to considering  self-adjoint  linear operators  $\K$, $\B$ and $\G$ in an appropriate Hilbert space, which $\Ps$ belongs to. Moreover, $\B$ and $\G$ and $\G^{-1}$ are assumed to be bounded.  We  keep two independent parameters $\delta$ and $\h$ in the next section, but it is assumed  later that $\delta=\sqrt\h$.

Depending on the relative complexity of the eigenvalue problems for the complete operator $\hK$ and the original one,  $\K$,  there are two ways of constructing the adiabatic expansion, which we present in the following two subsections.


\subsection{Adiabatic expansion in terms of perturbed $\hK=\hK(x,\delta)$}
We start by constructing the expansion in terms of the complete operator  $\hK$ (i.e., including the perturbation). Henceforth we denote the quantities corresponding to the complete operator $\hK$  with a hat.

We search for an adiabatic expansion of \Ref{mainEq} in the form of
\be\label{S-anzz}
	\hPs(x,\delta,\h)  =\hPh(x,\delta,\h)  \,e^{\tfrac{i}\h\int^x\hvt(x',\delta,\h)  \,dx'},
\ee
where  both $\hPh$ and $\hvt$ are given by formal series in powers of $\h$
\begin{eqnarray}
\hPh(x,\delta,\h) &=& \hPhn0{}{}(x,\delta) + \h\hPhn1{}{}(x,\delta) + \h^2\hPhn2{}{}(x,\delta) + \ldots,\qquad\\
\hvt(x,\delta,\h) &=& \hvtn0{}(x,\delta) + \h\hvtn1{}(x,\delta)\,\, + \h^2\hvtn2{}(x,\delta) +\ldots
		\label{S-anzatz1}
\end{eqnarray}
The standard adiabatic approximation contains only an expansion of $\Ph$, while we also introduce the second expansion in the phase factor. In a complete analogy with perturbation theory for eigenfunctions
(see \Ref{add-cond}), we are entitled to impose an additional condition
\be
	(\hPhn0{}{},\G\hPhn{n}{})=0, \qquad n\geq1.
\label{S-add_cond}
\ee
It fixes the arbitrariness of possible multiplication of the principal approximation by an arbitrary series in $\h$.
This condition makes the representation (\ref{S-anzz})--(\ref{S-anzatz1}) unique. As we will see in the sequel, it also guarantees that the amplitude factor $\hPh$ depends  on the local properties of the medium only, while all the integral (nonlocal) ones are contained in the phase factor.

Inserting (\ref{S-anzatz1}) into (\ref{mainEq}) and equating coefficients at equal powers of $\h$, we obtain a sequence of equations
\begin{eqnarray}
	(\hK-\hvtn0{}\G){\hPh}^{(0)}&=&
		0,\label{0-th}\\
	(\hK-\hvtn0{}\G){\hPh}^{(1)}&=&
		 \hvtn1{}\G{\hPh}^{(0)} -i\G\partial_x{\hPh}^{(0)},
		 \label{2-nd}\\
	\ldots&&\nonumber\\
	(\hK-\hvtn0{}\G){\hPh}^{(n)}&=&
		\hvtn{n}{}\G{\hPh}^{(0)}+\sum_{i=1}^{n-1}\hvtn{i}{}\G{\hPh}^{(n-i)}
		-i\G\partial_x {\hPh}^{(n-1)}.
		\label{n-th}
\end{eqnarray}
These equations differ from  known in quantum mechanics,   see \cite{qm_landavshitz_59},  by the presence of the matrix $\G$ and by the expansion in the exponent.

 To construct an adiabatic mode, we shall choose the principal approximation as
\begin{equation}
	\hPhn{0}{1} = \vvph_1, \qquad \hvtn0{1}=\hb_1,
\end{equation}
where $\hb_1$ and $\vvph_1$ are a real eigenvalue and the corresponding eigenfunction, respectively, of the perturbed spectral problem
\be
	\hK\vvph \equiv (\K+\delta \B)\vvph = \hb\G\vvph.
	\label{K-EVP}
\ee
Their properties, in particular, the orthogonality of eigenfunctions,  are discussed in Appendix \ref{app:eig-pr}.  Henceforth we supply the approximations ${\hPh}^{(n)}$ and ${\hvt}^{(n)}$ with a subscript, which reflects the indexing number of the eigenvalue and eigenfunction used in the principal approximation.

Solving the equations one by one, as the standard perturbation theory prescribes,
 in the principal approximation  we obtain
\begin{eqnarray}
&&	\hPs_1 = \hPs_1^{(0)} + O(\h),  \nonumber\\
&&  \hPs_1^{(0)} =
{\vvph_1}\exp{\frac{i}\h \int\limits^x_{x^*} \, \hb_1(x',\delta) dx' - \int\limits^x_{x^*} \hat S_{\brr11}(x',\delta) dx' }
, 	\label{hWKB1}
\end{eqnarray}
here
\be
	\hat S_{ij}\equiv \frac{(\vvph_i,\G\partial_x\vvph_j)}{\hat N_i},\qquad
\ee
$\hat N_1=({\vvph_1}, \G {\vvph_1}) $, and $x^*$ is a constant.  The second term in the exponent in \Ref{hWKB1} has a two-fold interpretation. Its  imaginary part is the Berry phase \cite{Berry-phase}, while its real part fixes the normalization of the whole solution. Indeed, analyzing the real part of $\hat S_{11}$ for smooth  $\hat N_1=\hat N_1(x,\delta)$, we find
\be
\Re \(\hat S_{\brr11}(x')\)
	= \frac{(\partial_x \vvph_1, \G \vvph_1)	
			+(\vvph_1,\G \partial_x \vvph_1) }
		{2(\vvph_1,\G  \vvph_1)}
	=  \frac{1}{2} \partial_x \ln {|\vvph_1, \G \vvph_1|},
	\label{Re S11}
\ee
and thus upon integration and exponentiation, as in \Ref{hWKB1}, at the upper limit of integration it gives exactly $|\hat N_1(x,\delta)|^{-\tfrac12}$ and a constant at the lower one, so that
 \be
\hPs_1^{(0)} = |\hat N_1(x^*,\delta)|^{\tfrac12}
\frac{\vvph_1(x,\delta)}{|\hat N_1(x,\delta)|^{\tfrac12} }\exp{\frac{i}\h \int\limits^x_{x^*} \, \hb_1(x',\delta) dx' - i \int\limits^x_{x^*}\Im{ \hat S_{\brr11}(x',\delta) } dx' }
, 	\label{hWKB1-norm}
\ee

 \be
	(\hPs_1, \G \hPs_1) = |\hat N_1(x^*,\delta)|\frac{\hat N_1(x,\delta)}{|\hat N_1(x,\delta)| } + O(\h).
\ee
Thus, the adiabatic mode can be made $\G$-normalized, $|(\hPs_1, \G \hPs_1)|=1+ O(\h)$, assuming the eigenfunctions are normalized in $x^*$. The overall sign of the normalization factor  $(\vvph_1,\G \vvph_1)$, however, cannot be fixed.
Similarly to \Ref{Re S11}, the presence of the Berry phase, i.e., $\Im \hat S_{\brr11}$,  makes \Ref{hWKB1} invariant under nonconstant  phase shifts of the eigenfunction, $\vvph_j\to e^{i\sigma_j(x)}\vvph_j$, with arbitrary smooth $\sigma_j(x)$. Thus, the adiabatic solution is uniquely defined up to a constant factor.

In the present paper, we are interested in the case $\delta=\sh$. Then,
 \Ref{hWKB1} contains an overmatch of accuracy.  Moreover, solving the eigenproblem \Ref{K-EVP} for $\hK=\K+\sh \B$ can be much more complicated then solving it just for $\K$.

As mentioned above, there are two complementary ways of constructing the adiabatic modes in the latter case. One could be the substitution into  \Ref{hWKB1} for $\delta=\sh$ of an expansion of the $\vvph$ and $\hb$ in powers of $\sh$ obtained by the perturbation method in Section \ref{sec:EVP-away}. This procedure involves the substitution of one asymptotic expansion into another, and we prefer to use an alternative, more transparent method.
It consists in building a separate perturbation scheme for the adiabatic mode itself. We give it in full details in the next subsection. The results of both ways do coincide,   as  will be clear from what follows.

\subsection{Adiabatic solution in terms of the original $\K$}\label{sec:wkb}
We search for adiabatic solution of \Ref{mainEq} in the same form
\be\label{anzz}
	\Ps (x,\sh) =\Ph(x,\sh) \,e^{\tfrac{i}\sh\int^x\vt(x',\h)} \, dx',
\ee
but now, both $\Ph$ and $\vt$ are given by formal series not in powers of $\h$, but of  $\sh$, respecting the order of magnitude of the perturbation
\begin{eqnarray}
	\Ph(x,\sh)&=&\Phn0{}{}(x) +\sqrt\h\Phn{1}{}{}(x)  + \h\Phn2{}{}(x)  + \ldots,
		\label{anzatz01}\\
	\vt(x,\sh)&=&\vtn0{} + \sqrt\h\vtn{1}{}(x)  + \h\vtn2{}(x)  +\ldots
		\label{anzatz1}
\end{eqnarray}
Similar to \Ref{S-add_cond}, we supply the representation (\ref{anzz}--\ref{anzatz1}) with the additional condition
\be
	(\Phn0{}{},\G\Phn{n }{})=0, \qquad n\geq1,
\label{add_cond}
\ee
which has the same meaning as \Ref{S-add_cond}. We also note that in the absence of the perturbation $\B$, both expansions (\ref{anzatz01}), (\ref{anzatz1}) are just those obtained in the previous section.

Following the perturbation method, we insert  (\ref{anzatz01}),(\ref{anzatz1}) into (\ref{mainEq}).  Equating the coefficients at equal powers of $\sqrt\h$, we obtain a  sequence of equations
\begin{eqnarray}
	(\K-\vtn0{}\G){\Ph}^{(0)}&=&
		0,\label{0-th}\\
	(\K-\vtn0{}\G){\Ph}^{(1)}&=&
		 -\B{\Ph}^{(0)}+\vtn{1}{}\G{\Ph}^{(0)},
		\label{1-th} \\
	(\K-\vtn0{}\G){\Ph}^{(2)}&=&
		 -\B{\Ph}^{(2)}+\vtn2{}\G{\Ph}^{(0)}+\vtn{1}{}\G{\Ph}^{(1)}-i\G\partial_x{\Ph}^{(0)},
		 \label{2-nd}\\
	\ldots&&\nonumber\\
	(\K-\vtn0{}\G){\Ph}^{(n)}&=&
		-\B{\Ph}^{(n-1)}+\vtn{n}{}\G{\Ph}^{(0)}\nonumber\\
			&&+\sum_{i=1}^{n-1}\vtn{i}{}\G{\Ph}^{(n-i)}
		-i\G\partial_x {\Ph}^{(n-2)}.
		\label{n-th}
\end{eqnarray}

Aiming at constructing  the `first' mode, we choose  as the solution of the principal  order equation (\ref{0-th}) the first eigenfunction of $\K$
\begin{equation}
	\Phn{0}{1} =  \vph_1, \qquad \vtn0{1}= \b_1,
	\label{Phn0}
\end{equation}
where $ \b_1$ and $ \vph_1$ are an eigenvalue and the corresponding eigenfunction, respectively, of the original spectral problem  \Ref{newEVP}.

We solve equations (\ref{1-th}), (\ref{2-nd}) and (\ref{n-th}) step by step.  All these equations
are solvable if their right-hand sides are orthogonal to the solution of the homogeneous equation (\ref{0-th}). This condition with account of (\ref{add_cond}) and \Ref{Phn0} yields
\begin{eqnarray}
&&\vtn1{1}=\frac{ \Bc_{\brr11}}{  N_1}, \quad
\vtn2{1}=\frac{( \vph_1, \B \Phn {1}{1}) +i ( \vph_1, \G \partial_x  \vph_1 )   }{   N_1}, \label{sec:t}\\
&&\vtn{n}{1}=\frac{( \vph_1, \B \Phn {n-1}{1}) +i ( \vph_1, \G \partial_x \Phn{n-2}{1} ) }{   N_1}, \quad  n>2, \label{b(n)}
\end{eqnarray}
We recall  here that the eigenvalues $ \b_j$, $j=1,2$,  are assumed real.

Taking into account (\ref{add_cond}), we write  the higher order approximations in the form
\be\label{first-ap}
	\Phn{n}{1}=\cn{n}{12} \vph_2 +  \vph_{1\perp}^{(n)},
	\qquad n\geq 1,
\ee
where $\cn{n}{12}$ is a scalar function of $x$, and  $ \vph_{1\perp}^{(n)}$ is $\G$-orthogonal to $ \vph_1$ and $ \vph_2$
\be\label{ort-def}
	( \vph_j, \G \vph_{1\perp}^{(n)})=0, \qquad j=1,2.
\ee
In what follows, we shall consider the degeneracy between $\b_1$ and $\b_2$, so we separate the  term proportional to $\vph_2$, because it contains the main singularity when  $\b_2$ is close to $\b_1$, see Section \ref{sec:wkb-valid}.
To find $\Phn{n}1$, we substitute (\ref{first-ap}) into (\ref{n-th}), calculate its scalar product with  $\vph_2$, taking into account (\ref{ort-def}) and the orthogonality properties of eigenfunctions, see Appendix \ref{app:eig-pr}. We find
\begin{eqnarray}
&&\cn{1}{12}=\frac{ \Bc_{\brr21}}{( \b_1- \b_2)  N_2},
		\label{c-12-1}\\
&&\cn{n}{12}
	=\frac{( \vph_2, \B \Phn {n-1}{1}) - N_2 \sum_{i=1}^{n-1}\vtn{i}{1} \cn{n-i}{12}
		+ i ( \vph_2,\G \partial_x \Phn{n-2}{1} ) }{ ( \b_1- \b_2)  N_2}, n\ge 2.
					 \label{c-12-n}
\end{eqnarray}

To complete the construction of the adiabatic solution, we need to deduce the perpendicular component $\vph_{1\perp}^{(n)}$, $n=1,2,\ldots$ However, as we shall see in the next section, it does not influence the transition of the modes, at least in the principal order. Thus, we only need to show the possibility of its determination to prove that the recurrent system (\ref{0-th})-(\ref{n-th}) can indeed be solved  step by step. We rewrite equation (\ref{n-th})  as follows
\be\label{def-vperp}
	(\K-\b_1\G)\vph_{1\perp}^{(n)} =
		 \G \f^{(n)},
\ee
where $\G\f^{(n)}$ is the right-hand side of (\ref{n-th}).
 In view  of  the assumption of the presence of a finite gap between $\b_j,$ $j=1,2$ and the rest of the spectrum and the fact that $\f^{(n)}$ is $\G$ -- orthogonal to $\vph_j$, $j=1,2$, equation \Ref{def-vperp} has a single solution $\vph_{1\perp}^{(n)}$; see  Appendix A, (property 5).

In the case of a purely discrete spectrum of $\K$, the first order approximation of $\vph_{1\perp}^{(n)}$ has the  form
\be
	\vph_{1\perp}^{(1)}= \sum\limits_{j\ne1,2} c_{1j}^{(1)} \vph_j = \sum\limits_{j\ne1,2} \frac{ \Bc_{\brr j1}}{( \b_1- \b_j)  N_j} \vph_j,
\ee
where the summation ranges  all  the modes except for the first two.
Thus, all the terms of the formal series  can be constructed.

We assume now that $|N_1(x^*)|=1.$ In the $O(\h)$ approximation, we thereof have
\be
\Ps_1= \frac{1}{|N_1(x)|^{1/2}}  \( \vph_1
	+\sqrt\h\frac{ \Bc_{\brr21}}{( \b_1- \b_2)   N_2} \vph_2
	+\sh \vph_{1\perp}^{(1)}
	+\ldots\) \times
		\label{WKB1}
\ee
$$
\exp{i\Theta_1+\frac{i}\h\int^x_{x^*} \vartheta_1 dx},
$$
where  we employ the  notation
\be
\vartheta_1=
\b_1+\sqrt\h \frac{ \Bc_{\brr11}}{  N_1}
 +\h\[\frac{  \Bc_{\brr21}  \Bc_{\brr12}}{ ( \b_1- \b_2)   N_1  N_2 }
	+\frac{ ( \vph_1,\B  \vph_{1\perp}^{(1)}) }{N_1}- {\rm Im}   S_{\brr11} \]
	+O(\h^{3/2})
\ee
and
\be \label{def-B-perp}
	  S_{ij}\equiv \frac{( \vph_i,\G\partial_x \vph_j)}{N_i},
\ee
$\Theta_1$ is a constant phase factor.
To obtain $\vtn{2}{1}$ explicitly we have used (\ref{first-ap}), (\ref{c-12-1}), the orthogonality condition for $\vph_j$, and the notation (\ref{def-B-perp}).

First of all, we note that \Ref{hWKB1}, together with \Ref{vvph} and \Ref{hb}, gives \Ref{WKB1}, as expected.   Our comments on the treatment of  $\Im \hat S$ are equaly applicable to $\Im S$ in \Ref{WKB1}.
We note that, according to the property 3 of Appendix \ref{app:eig-pr}, the normalization factor $N_j(x)$ is not equal to zero. Its sign cannot be fixed, and represents intrinsic properties of solution. Formula \Ref{WKB1} also shows that the whole adiabatic solution $\Ps_j$ is normalized in the principal approximation.

Recall that the structure of the amplitude $\Ph$ guarantees that under the transformation $\vph_j\to e^{i\sigma_j}\vph_j$, $j=1,2$ (no hats!),  it maps in the same way: $\Ph_j\to e^{i\sigma_j}\Ph_j$. And at the same time, the Berry phase $\int^x {\rm Im}   S_{\brr11} dx$ transforms nontrivially under the same phase shift if $\s_j$ is not constant, $\s_j=\s_j(x)$,
\be
	\int^x_{x^*} {\rm Im} S_{\brr11} dx  \mathop\to_{\vph_j\to e^{i\sigma_j}\vph_j}
		\s_1(x) - \s_1(x^*) +\int^x_{x^*} {\rm Im}   S_{\brr11} dx.
\ee
So, the only ambiguity left in the definition of $\Ps_j$ is an overall {\it constant} phase factor. It can be interpreted purely in terms of the lower limit of integration $x^*$, but  for simplicity of further analysis we introduced an additional parameter $\Theta_j$. It will be chosen in Section \ref{sec:can-modes} in such a way that the dependence on $x^*$ will disappear from  \Ref{WKB1} whatsoever.


We  call the formal series constructed here {\it the adiabatic expansion} or  {\it adiabatics}. The principal term of the expansion  is named the adiabatic approximation or adiabatic mode.

The other solution, $\Ps_2$ is obtained by simply interchanging the indices $1\leftrightarrow 2$.   The formula \Ref{WKB1} works if there is a finite gap between $\b_1$ and the rest of the spectrum. If the distance between $\b_1$ and $\b_2$ at a point decreases with decreasing of $\h$, adiabatic formulas may not work, as follows from \Ref{first-ap}, \Ref{c-12-1}, and \Ref{c-12-n}. We investigate in the next section the validity domain of \Ref{WKB1}.



\subsection{The validity region of adiabatic solutions and the slow variable}
\label{sec:wkb-valid}

As already mentioned in the Introduction, we consider in this paper only the case of the intersection of (real) eigenvalues of original spectral problem \Ref{newEVP}
\be
	\b_{2} - \b_1 \seq_{x\to0}  2 Q x, \qquad  Q>0.
	\label{Qx}
\ee
We show now that the higher-order terms both in the amplitude and in the exponent of the adiabatic solution \Ref{WKB1} have a singularity at the degeneracy point and the order of the singularity increases with the order of approximation. In other words, in the vicinity of the degeneracy point $x=0$, the adiabatic expansion looses its asymptotic character.

First of all we note that on an interval surrounding $x=0$ the matrix elements $ \Bc_{\brr jk}$ and $  S_{\brr jk}$, $j,k=1,2$, are bounded. This fact follows from
the assumption that $\B$, $\G$ are bounded and functions $ \vph_j$ and their $x$ derivatives can  always be chosen continuous and have a bounded norm $\| \vph_j\|<\infty$,  $\|d  \vph_j/dx\|<\infty$.
From  \Ref{def-vperp} it follows that
\be\label{v-perp-est}
\|\vph_{1\perp}^{(n)}\|
	= \|(\K- \b_1\G)^{-1}  \G \f^{(n)}\| \le \|(\G^{-1}\K- \b_1)^{-1}_\perp\|   \|  \f^{(n)}\|.
\ee
The restriction of   $(\G^{-1}\K- \b_1)^{-1}$ to the subspace $\G$ -- orthogonal to $\vph_j$, $j=1,2$,
denoted above by $(\G^{-1}\K-\b_1)^{-1}_\perp$, is bounded because both  $ \b_1$ and $ \b_2$ are separated from the rest of the spectrum, and because $\G$ and its inverse are bounded.  The right-hand side of equation (\ref{1-th}) denoted by $\G\f^{(1)}$ has a bounded norm, as follows from the above-said and the first formula of \Ref{sec:t}.  Therefore $\vph_{1\perp}^{(1)}$ does not have a singularity at $x=0$. The singularity of $\Phn{1}{1}$ is determined by the singularity $\cn{1}{12}$ and, taking into account (\ref{c-12-1}), we obtain
$\Phn{1}{1}= O\((\b_1-\b_2)^{-1}\).$  Formula  (\ref{sec:t}) shows that $\vtn{2}{1}$ is of order $\Phn{1}{1}$, $\vtn{2}{1}= O\((\b_1-\b_2)^{-1}\)$.
The nominator of $\cn{2}{12}$, according to (\ref{c-12-n}) for $n=2$, is of order $\cn{1}{12}$; note that $\Phn{1}{1} \sim \cn{1}{12}$.  The difference of the eigenvalues $\b_1$ and $\b_2$ in the denominator yields $\cn{2}{12}= O\((\b_1-\b_2)^{-2}\).$
To estimate $\vph_{1\perp}^{(2)}$, we note that the right-hand side of (\ref{2-nd}) is of order $O\((\b_1-\b_2)^{-1}\)$. Therefore
$\vph_{1\perp}^{(2)} = O\((\b_1-\b_2)^{-1}\)$, according to (\ref{v-perp-est}). Thus we have $\Phn{2}{1}= O\((\b_1-\b_2)^{-2}\).$
To estimate $\vtn{3}{1}$ from (\ref{b(n)}) we need also estimate $\partial_x \Phn{1}{1},$ which is of order  $d \cn{1}{12}/dx = O\((\b_1-\b_2)^{-2}\)$, because $d({ \b}_1 - { \b}_2)/dx$ is a constant.

The process can be continued. The singularity of  $\Phn{n}{1}$ is the same as the singularity of $\cn{n}{12}$, which is calculated by (\ref{c-12-n}). It is given by a fraction, where all terms in the nominator are  of order $O\((\b_1-\b_2)^{-(n-1)}\)$ as $\cn{n-1}{12}$ and $d\cn{n-2}{12}/dx.$
The denominator contains $\b_1-\b_2$. The order of the singularity of $\vph_{1\perp}^{(n)}$ is smaller.
The singularity of $\vtn{n}{1}$ is of order $\cn{n-1}{12}$.
Finally, we obtain
\be
\begin{aligned}
	\Phn{n}{1} \sim \cn{n}{12}\vph_2 = O\( {(\b_1-\b_2)^{- n }}\) = O(x^{-n}),\\
	\vtn{n}{1}= O\( {(\b_1-\b_2)^{- n+1 }}\) = O\( {x^{ -n+1}}\)
\end{aligned}
\ee
near $x=0$.

Then the $n$--th correction to the adiabatic expansion (\ref{WKB1})  behaves as
$$
	\h^{n/2} \Phn{n}\, = \,	O\(  \frac{\h^{n/2}}{x^{n}}\).
$$
It stays small and thus guarantees the asymptotic nature of expansions \Ref{WKB1} for
\be
	x\sim \h^{1/2-\gamma}\gg \h^{1/2}
\label{wkb_zone}
\ee
for any $\gamma>0$. This suggests to seek the \emph{resonant} or \emph{inner} expansion of (\ref{mainEq}) in the vicinity of $x=0$ in terms of the slow, or stretched,  variable
\be
	\tau=x/\sqrt\h,
	\label{slVar}
\ee
and in terms of this variable we shall re-expand the adiabatic expansion for future matching with the inner one.


\section{Transition matrix and numbering of modes}\label{sec:numbering}

As shown above, the adiabatic solutions $\Ps_j$, $j=1,2$ cannot be defined on an interval containing a degeneracy point. The same applies to $\hPs_j$, $j=1,2$, at least for not too big $|\nu|$, see Appendix \ref{app:outer-lift}.  Establishing a connection between the adiabatic modes on two sides of a degeneracy point is the essence of the connection problem considered in the present paper.

Let two eigenvalues $\b_1$ and $\b_2$ be separated from the rest of the spectrum with a gap. Let the asymptotics as $\h \to 0$ of an exact solution $\Om$ be a linear combination of the corresponding adiabatic modes on one side of the degeneracy point, say, to its left, $x< 0$,
\be
	\Om\seq_{x \ll -\sh}  k_1^- \Ps_1 + k_2^- \Ps_2;	
\ee
 then the aim of the \emph{connection problem} is to find them on the other side, for $x> 0 $,
\be
	\Om\seq_{x \gg \sh} k_1^+ \Ps_1 + k_2^+ \Ps_2.
\ee
By the adiabatic theorem, the discrepancy is maximum of order $O(\h)$.
This problem is solved by calculating the transition matrix connecting the two sets of coefficients
\be
 \T \(\begin{array}{c}
		 k_1^-\\
		 k_2^-
		\end{array}\)
=	\(\begin{array}{c}
		 k_1^+\\
		 k_2^+
		\end{array}\).
\label{trMa}
\ee
 For an unambiguous definition of $\T$ one should fix the normalization and the phases of the modes $\Ps$, which we carry out in  Section \ref{sec:can-matr}, and also their numbering.

Generally speaking, since the adiabatic modes are defined separately on different sides of a degeneracy point,  their numbering (i.e., the choice of the index $j=1,2$) is a question of our convenience, and there are two ways of doing it. Each of the numbering methods has its benefits and implications.

Throughout this paper, we fix the numbering of adiabatic solutions, $\Ps_j$, $j=1,2$,  of the original operator  by the condition \Ref{mb-cross}. This choice of numbering is based on the smoothness of the eigenvalues $\b_i$ and on the sign of the derivative of $\b$ at the degeneracy point. It is most convenient for $N_1N_2<0$. In this case, the sign of the flux of each mode, ${\sgn} N_i$, is maintained across the degeneracy point,  because $N_i(x)\ne 0$, see  Appendix \ref{app:eig-pr}, and it is a smooth function of $x$ if corresponds  to smooth eigenvalues. In subsection \ref{sec:can-matr}, we show that as $\nu\to 0$  the    transition between $\Ps_j$, $j=1,2$  is govern by a diagonal matrix.
To have the identity transition matrix, one needs to fix correctly the constant factors between modes, (see  Section \ref{sec:can-matr}).

However, this choice is not the most natural one for an avoided crossing.
Indeed, in this case a perturbed eigenvalue approximating an original one with a given number becomes a nonsmooth function, as  follows from considerations of Section \ref{sec:EVP-close}.
A more appropriate way would be the numbering that follows the sign of eigenvalues, not that of their derivatives. For details, see formulas \Ref{hb-b-t+}, \Ref{hb-b-t-} and Figs. \ref{real_TP}, \ref{compl_TP}.

As was mentioned, we follow the numbering \Ref{mb-cross}, more appropriate for the case of real degeneracy points. The transition matrix for the other numbering is given by  \Ref{number-T-LZ}.


\section{Inner asymptotic expansion}\label{sec:inner}

Adiabatic formulas fail to hold near $x=0$, as we showed in Section \ref{sec:wkb-valid}. To match adiabatic modes at different sides of the degeneracy point, $x=0$,  we construct here a formal asymptotic expansion valid in its vicinity and name it \emph{the inner} or \emph{ resonance expansion}.  To do this, we first express \Ref{mainEq} in terms of the slow variable $\tau$. Inserting expansions \Ref{expan-H-B} of $\K (\sh \tau)$ and $\B (\sh \tau)$  into  \Ref{mainEq}, we obtain an equation
\be
	\(\Kn0+\sh (\tau \Kn1 +  \Bn0)
		+\h(\tau^2\Kn2 + \tau\Bn1)
			+ \ldots\)\ps=-i\sh\G\frac{\partial\ps}{\partial \t}.
	\label{resoEq}
\ee
Its solution can be sought in the form of
\be
	\ps = \ph \,e^{\tfrac{i}\sh\int\limits^\t_{-b} \(\b_0 + \sh \hb_{av}^{(1)}\) d\t'},\quad  \ph=\phn0+\sh\phn1+\h\phn2+\ldots,
    \label{anzatz}
\ee
where $\b_0$ and $\hb_{av}^{(1)}$ are defined in \Ref{b-0-def} and \Ref{b-av}.
Substituting \Ref{anzatz} into \Ref{resoEq} and equating the coefficients at equal powers of $\sh$, we obtain a sequence of equations
\begin{eqnarray}
(\Kn0-\b_0\G)\phn0&=&		0\label{0-th-res}\\
(\Kn0-\b_0\G)\phn1&=&	(\hb_{av}^{(1)}\G - \tau \Kn1 - \Bn0)\phn0  -i\G\dot{\ph}^{(0)}, \label{st1-th-res} \\		
(\Kn0-\b_0\G)\phn2&=&   (\hb_{av}^{(1)}\G -  \tau\Kn1 - \Bn0)\phn1   \nonumber\\
&& -(\tau^2\Kn2 + \tau\Bn1)\phn0 -i\G\dot{\ph}^{(1)}, \label{st2-nd-res}\\
	\ldots &&\nonumber
\end{eqnarray}
where the dot marks the derivative with respect to $\tau$, $\dot{f}\equiv \partial f/\partial  \tau$.
 We note that the equations \Ref{0-th-tau}--\Ref{2-nd-tau} are transformed into \Ref{0-th-res}--\Ref{st2-nd-res} upon replacing $\hbn0$, ${\vph}^{(0)}$ $\vvphn1$  and $\hbn1$ by  $\b_0$, $\phn0$, $\phn1$, and $\hb_{av}^{(1)}-i \partial /\partial  \tau$, respectively. Thus we can repeat the  line of argument of Subsection \ref{sec:EVP-close}, also replacing $\a_{jk}^{(0)}(\tau)$ by
$a_{k}^{(0)}(\tau),$ $k=1,2$, to avoid a conflict of  notation.

Thus, the principal approximation of the amplitude, $\phn0$,  is given by the  formula
\be
    \phn0 = a_1(\tau)  \vph_1(0) + a_2(\tau) \vph_2(0),
	\label{phi0}
\ee
where the scalar coefficients $a_k(\tau), k=1,2$, satisfy a system of ordinary differential equations
\be
\begin{aligned}
	& 	(N_1^{(0)}\, (\,-i {\partial /\partial  \tau} + \hb_{av}^{(1)}\,)\,  - \tau \Kcn1_{11} - \Bcn0_{11})\, a_{1}
		+ (-\tau \Kcn1_{12} - \Bcn0_{12}) \, a_{2}=0\\
	&   (-\tau \Kcn1_{21}- \Bcn0_{21}) \, a_{1}
	+(N_2^{(0)}\,(\,-i  {\partial /\partial  \tau} + \hb_{av}^{(1)}) - \tau \Kcn1_{22} - \Bcn0_{22}) \, a_{2}=0.
\end{aligned}
\ee
 Using the definitions \Ref{b-av}, \Ref{not-Q} and \Ref{not-b},  it can be written
\be
\begin{aligned}
	& 	-i \dot a_1 =
		-(\tau +b)Q   a_1 		+   \frac{B_{12}^{(0)}}{N_1^{(0)}}   a_2,
	\\
	&	-i \dot a_2 =
		\frac{B_{21}^{(0)}}{N_2^{(0)}} a_1 		+(\tau +b) Q  a_2 .
\end{aligned}
\ee
%
To solve this system, we first express $a_1$ from the second equation in the form
\be
	a_1= -\frac{N_2^{(0)}}{B_{21}^{(0)}}\, \(i \dot a_2 + (\tau +b) Q a_2\) 
	\label{a1}
\ee
and substitute it into the first one to obtain the following equation for $a_2$:
\be
	 \ddot a_2  + \((\tau +b)^2 Q^2 - i Q + \nu \s^2 \)a_2  =0; 
	\label{a2}
\ee
here
\be
	\s =    e^{-i\pi/4}\sqrt{2Q},
	\label{sigma}
\ee
 and $\nu$ is given by \Ref{nu_gen}.

We note that in terms of a new variable
\be
	t = \s(\tau +  b ), \label{def-t-sig}
\ee
the equation (\ref{a2}) reduces to the parabolic cylinder equation \cite{abramowitz}
\be
	\partial^2_t y(t)
	+ \(\frac12-\frac{t^2}4+\nu\) y(t)=0,
		\label{ParCyl}
\ee
where $a_2(\tau)= y(\s(\tau +  b ))$, the parameter $\nu$ was introduced in \Ref{nu_gen}.
The general solution to (\ref{ParCyl}) can be written as a linear combination
\be\label{a_2t}
	a_2(\tau) \equiv y(\s (\tau + b) )= A D_\nu(\s(\tau +  b ))+B D_\nu (-\s(\tau +  b )),
\ee
where $A$  and $B$ are arbitrary constants, and $D_\nu$ is the parabolic cylinder function in the notation of Whittaker \cite{Whittaker}. By using its property \cite{abramowitz}
$$
	\partial_t D_\nu(t)+\frac{t}2 D_\nu(t)-\nu D_{\nu-1}(t)=0,
$$
from (\ref{a1})  we derive the expression for the  first  coefficient in  \Ref{phi0}
\be\label{a_1t}
	a_1(\tau)= - i \frac{B_{12}^{(0)}}{ \s N_1^{(0)}}\,\, \( A D_{\nu-1}(\s(\tau +  b ))-B D_{\nu-1} (-\s(\tau +  b ))\),
\ee
where we take into account that
\be
\frac{\nu \s N_2^{(0)}}{\Bc_{21}^{(0)}} = \frac{\Bc_{12}^{(0)}}{ \s N_1^{(0)}}.
\ee

Using the formulas \Ref{anzatz}, \Ref{phi0}, \Ref{a_2t}, \Ref{a_1t}  and $\s$ from  \Ref{sigma}, we obtain the principal approximation of the inner expansion
\be
	\ps(\tau,\h) =  \phn0(\tau) \,e^{\tfrac{i}\sh\int\limits^\t_{-b} \(\b_0 + \sh \hb_{av}^{(1)}\) d\t'}  + O( \sh ),
\label{conn-ps}
\ee
\be
\phn0(\tau)  = -\frac{i \Bc_{12}^{(0)}}{\s N_1^{(0)}}
	\[A D_{\nu-1}(t)-B D_{\nu-1} (-t)\]  \vph_1(0) 
	+ \[ A D_\nu(t)+B D_\nu (-t)\]  \vph_2(0), \label{conn-ph0}
\ee
$$  t = \s (\tau +  b ). $$

The process of constructing inner solution can be continued. One can show that in higher approximations, the relation
\be
	\h^{n/2} \phn{n} =  O(\h^{n/2} \tau^{2n+1})
\ee
holds. Then the validity region of the inner solution is given by
\be
	|\tau| \sim \h^{-1/4 +\g'}, \quad  |x| \sim \h^{-1/4+\g'}
\ee
with arbitrary small $\g'>0$. Comparing with the validity region of the adiabatic solution \Ref{wkb_zone}, we observe that the intersection of these  regions or the \emph{matching region} is
\be
	|x| \sim \h^{1/2 - \g}, \quad 0 <\g< 1/4.
  \label{inters-zone}
\ee

\section{Inner expansion in the matching region}\label{sec:inner-matching}
Following the matched asymptotic expansion method, we derive now the principal term of the inner  expansion of $\ps$ given by \Ref{conn-ps}, \Ref{conn-ph0} in the matching region, i.e., on the outskirts of the validity domain, as $\tau\equiv x/\sh \to \pm \infty$, but \Ref{inters-zone} is still satisfied.

The asymptotics of  $D_\nu(t)$   is different in different sectors of the complex plane \cite{abramowitz}
\begin{eqnarray}
D_\nu(t) \seq_{|t|\to\infty} t^\nu e^{-\frac{t^2}{4}}(1+O(t^{-1})),&\qquad\qquad
		&\arg(t)\in (-\frac{3\pi}{4}, \frac{3\pi}{4}),  \label{Dnu2}\\
D_\nu(t) \seq_{|t|\to\infty} t^\nu e^{-\frac{t^2}{4}}(1+O(t^{-1}))&
		+ \xi_\nu e^{2 i\pi\nu}& t^{-\nu-1} e^{ \frac{t^2}{4}}(1+O(t^{-1})),\nonumber \\
&\quad			&\arg(t)\in (\frac{\pi}{4}, \frac{5\pi}{4}),
\label{Dnu3}
\end{eqnarray}
where
$$
	\xi_\nu \equiv -\frac{\sqrt{2\pi}}{\Gamma(-\nu)} e^{-i\pi\nu},
$$
for the future use we note that $ \xi_{\nu-1}= \xi_\nu /\nu$.


To calculate the asymptotics of \Ref{a_1t}, \Ref{a_2t}, we need the following observation
\be\label{D-1}
D_{\nu}(\pm\s (\t +b)) = e^{i\nu\,\arg{(\pm\s\t)}}\,   |\s\t |^{\nu} e^{-\frac{\s^2 (\t+b)^2 }{4}}
	+ O(\t^{-1})
\ee
valid for $\arg{(\pm \s \t)} = -\pi/4$, or $\arg{(\pm\s \t)} = 3\pi/4$ and $|\t| \to \infty$. Here $b$ is a fixed   constant. We also took into account the fact that for imaginary $\nu$, the second term in \Ref{Dnu3} is of order $t^{-1}\sim \tau^{-1}$, while the exponent factor is a pure phase, $|\exp{\pm t^2/4}|=1$.

We need also the  asymptotics
\be\label{D-2}
D_{\nu-1}(\pm\s (\t +b)) = O(\t^{-1}),
\ee
valid for $\arg{(\pm \s \t)} = -\pi/4$, $|\t| \to \infty$ and
\be\label{D-3}
D_{\nu-1}(\pm\s (\t +b)) = e^{-i\nu\arg{(\pm\s\t)}}\, \xi_{\nu-1} e^{2 i\pi\nu} |\s\t |^{-\nu} e^{\frac{\s^2 (\t+b)^2 }{4}} + O(\t^{-1})
\ee
if $\arg{(\pm \s \t)} = 3\pi/4$, $|\t| \to \infty$.

 We proceed now with the asymptotics to the left of the turning point. We fix the arguments, so that for negative $\t$ and $|\t|>|b|$,
$$
\arg(\s(\tau+b)) = \arg (e^{-i{\pi}/{4}} \tau)= 3{\pi}/{4},\qquad
\arg (-\s(\tau+b))=-\pi/4.
$$
Taking into account \Ref{D-1}-\Ref{D-3}, for the amplitude of the principal term of the inner expansion \Ref{conn-ph0} as
${\t \to -\infty}$  we obtain
\begin{equation}\label{neg-as}
\begin{aligned}
\phn0 \=_{\t \to -\infty}
	-  \frac{ i \xi_{\nu-1} \Bc_{12}^{(0)}}{\s N_1^{(0)}}\,    &A e^{i\frac{5\pi\nu}{4}}
 			\frac{ e^{\frac{\s^2(\t+b)^2}{4}}}{|\s\tau|^{\nu}}
			 \vph_1(0)& \\
			 +
			 &\(A e^{i\frac{3\pi\nu}{4}}   +  B e^{-i\frac{\pi\nu}{4}}\)
			 \frac {|\s\tau|^{\nu}}{ e^{\frac{\s^2(\tau+b)^2}{4}}}
				 \vph_2(0)  + O(\t^{-1}).&
\end{aligned}
\end{equation}

Let us turn now to the asymptotics to the right of the turning point. For $\t>0$ we put
$$
\arg(\s(\tau+b)) = - \frac{\pi}{4},\qquad
\arg (-\s(\tau+b))=3\pi/4.
$$
Formulas \Ref{D-1}-\Ref{D-3} in this case yield an asymptotics as ${\t \to +\infty}$
\be
\begin{aligned}
\phn0 \=_{\t \to +\infty}
	 \frac{  i \xi_{\nu-1}\Bc_{12}^{(0)}}{ \s N_1^{(0)}}  \,  &B e^{i\frac{5\pi\nu}{4}}
			 \frac{ e^{\frac{\s^2(\t+b)^2}{4}}}{|\s\tau|^{\nu}}
			 \vph_1(0)& \\
 + & \(A e^{-i\frac{\pi\nu}{4}}   +  B e^{i\frac{3\pi\nu}{4}}\)
			\frac {|\s\tau|^{\nu}}{ e^{\frac{\s^2(\tau+b)^2}{4}}}
				 \vph_2(0)  + O(\t^{-1}).&
\label{pos-as}				
\end{aligned}
\ee
In both cases,  \Ref{conn-ps} should be applied for the inner expansion.
\section{Adiabatic modes rearrangement in the matching region}
\label{adia_rearr}
We shall construct now the rearrangement of the adiabatic expansion on the boundary of its validity region with the above formulas.

First of all, formula \Ref{b-non-pert}  in terms of the slow variable $\tau$ \Ref{slVar} yields
\be\label{del-b}
	 \b_2 -  \b_1 =   2Q \sh \tau + O(\h \tau^2).
\ee
Taking into account the notation \Ref{b-0-def}, \Ref{not-Q},
  \Ref{not-b}, \Ref{nu_gen}, \Ref{b-av} and also \Ref{del-b}  and the definition of the slow variable \Ref{slVar},   in the exponent of \Ref{WKB1} we obtain
\begin{eqnarray}
&&
i\Theta_1 + \frac{i}{\h}\int^x_{x^*} dx' \vartheta_1(x', \h)
\nonumber\\
&&
= i\Theta_1 + \frac{i}{\h}\int^x_{x^*} dx'
	\(
	 \b_1
	+\sqrt\h \frac{ \Bc_{\brr11}}{  N_1}
	+ \h\[\frac{  \Bc_{\brr21}  \Bc_{\brr12}}{ ( \b_1- \b_2)   N_1  N_2 }
		+ \frac{  ( \vph_1,\B  \vph_{1\perp}^{(1)})}{  N_1} -{\rm Im}S_{\brr11} \]
	+ \ldots
	\)  \nonumber
\\
&&= i\Theta_1 +i \int^{\tau}_{\tau^*} d\tau'
	\( \frac{\b_0}{\sh} + \hb_{av}^{(1)}(\t') -   Q (\t' + b) +i \frac {\nu } {\tau' } + {\cal R}_1 \) \,  \,
\nonumber \\
&& = \frac{i}\sh\int\limits^\t_{{-b}}
	\(\b_0 + \sh \hb_{av}^{(1)}(\t')\) d\t' - i\frac{Q (\t +b)^2}{2}-\nu \ln|\s \tau|   + {\cal R}   + i\zeta ,
	\label{rearr_ph}
\end{eqnarray}
where the remainder ${\cal R}_1= O(\sh)$  comprises  higher terms like
$\sh (\frac{  ( \vph_1,\B  \vph_{1\perp}^{(1)})}{  N_1}-{\rm Im} S_{\brr11})$ and also the discrepancy caused by the replacement of $ \b_1$ and $ \b_1- \b_2$ by their principal terms.
The variable $\tau$ may be as large as $O(\h^{-\gamma})$, $0<\gamma<1/2$, while the lower limit, $\tau^*\equiv x^*/\sh$, is assumed to be much less, $\t^*\ll \t$. Thus ${\cal R}=O(\h^{1/2-\gamma})$.
From the third line of formula \Ref{rearr_ph}, we notice that under the integral there are the first three terms of the expansion \Ref{lar-t-as} of the exact eigenvalue $\hat\b_1(x,\h)$ for positive $\tau+b$ (or $\hat\b_2(x,\h)$ for negative $\tau+b$) in vicinity of the degeneracy point. We will use this observation in Section \ref{sec:can-modes} for interpreting $\Theta_1$ and $x^*$.

In passing from the third line to the forth, we have chosen $\Theta_1$ in such a way that the dependence on the lower limit of integration $\tau^*$ in \Ref{rearr_ph} disappears completely
\be
	i \Theta_1= i\zeta + \frac{i}\sh\int^{\t^*}_{{-b}} \(\b_0 + \sh \hb_{av}^{(1)}(\t')\) d\t'
	- i \frac{Q (\t^* +b)^2}{2}  -  \nu \ln|\s \tau^*|.
\label{Theta1}
\ee
Here $\zeta$ envolves the remaining arbitrariness in the definition of the phase of adiabatic modes, which will be fixed in Secion \ref{sec:can-matr}. Such a choice of $\Theta_1$ also guarantees that the formulas for the first mode (and the second one) are the same both for  negative and positive $\t$.

Second, the amplitude factor  in \Ref{WKB1} under the same assumptions as above becomes
 \be\label{rearr_amp}
 \Ph=
	\frac{1}{|N^{(0)}_1|^{1/2}} \(\vph_1(0)-\frac{ \Bcn0_{21}}{2 Q N_2^{(0)}\tau}\vph_2(0) +\ldots\).
\ee
 As  follows from the discussion  in Section \ref{sec:wkb-valid}, the approximations of  order  $\h^{n/2}$ after transition to the slow variable $\t$   both in the phase and the amplitude  become of  order $O(\t^{-n})$. But we consider the case $\tau =O(\h^{-\gamma})$, for arbitrary small  $\gamma>0$.
Therefore the unaccounted terms in (\ref{rearr_ph}), (\ref{rearr_amp}) and the second term in the latter formula are negligibly small, and the principal term of the first mode with account of notation \Ref{sigma} reads
\be
\Ps_{1}^{(0)}(\sh\t,\h)
	=  \,e^{\tfrac{i}\sh\int\limits^\t_{ {-b}} \(\b_0 + \sh \hb_{av}^{(1)}\) d\t'}
		e^{i\zeta} \, |\s\t|^{-\nu} \, e^{\frac{ \s^2 (\t+b)^2}{4}}
		\frac{\vph_1 (0)}{|N^{(0)}_1|^{1/2}}
	+ O(\tau^{-1}).
\label{Ps1p}
\ee
Interchanging the indices $1\leftrightarrow 2$ (which in \Ref{rearr_ph} accounts for changing the overall sign in all terms but the first one), for the second mode  we get
\be
\Ps_{2}^{(0)}(\sh\t,\h)
	=   e^{\tfrac{i}\sh\int\limits^\t_{ {-b}}  ( \b_0 + \sh \hb_{av}^{(1)})  d\t'}
		e^{-i\zeta} \, |\s\t|^{\nu} e^{-\frac{\s^2 (\t+b)^2}{4}}
		\frac{\vph_2 (0)}{|N^{(0)}_2|^{1/2}}
	+ O(\tau^{-1}),
\label{Ps2p}
\ee
and on the different sides of the degeneracy point both modes are given by the same formula  ($j=1,2$)
\be
\Ps_{j-}^{(0)}(\sh\t, \h) = \Ps_{j}^{(0)}(\sh\t,\h)   {\rm\ if} \, \t<0;\quad
		\Ps_{j+}^{(0)}(\sh\t,\h) = \Ps_{j}^{(0)}(\sh\t,\h) {\rm\ if} \, \t>0.
\label{Ps-pm}
\ee
We recall that $\nu$ and $\s^2 (\t+b)^2$ are purely imaginary and the principal terms \Ref{Ps1p} and \Ref{Ps2p} are of order one in $\t$.



\section{Transition matrix }\label{sec:T}
\subsection{Matching of inner and outer expansions}\label{sec:match}
We are ready now to match inner and outer expansions thus obtaining the transition matrix.

Comparing \Ref{neg-as} and \Ref{pos-as} with \Ref{Ps1p} and  \Ref{Ps2p} correspondingly, we immediately recognize that
\begin{eqnarray}
 && {|N_2^{(0)}|^{-1/2}}	\psn0
	 \mathop{\simeq}_{\tau\to-\infty}
		 \co_1^- \Ps^{(0)}_{1- } +    \co_2^-\Ps^{(0)}_{2- },
	\label{ps-infty-}\\
&&	{|N_2^{(0)}|^{-1/2}} \psn0
	 \mathop{\simeq}_{\tau\to+\infty}
		 \co_1^+ \Ps^{(0)}_{1+ } +    \co_2^+\Ps^{(0)}_{2+ }, \label{ps-infty+}
\end{eqnarray}
where
\begin{eqnarray}
\co_1^-  \equiv  - i \sqrt{\nu} e^{i\theta_a} A \xi_{\nu-1}  e^{ i\frac{5\pi\nu}{4}}e^{-i\zeta} ,\qquad
\co_2^- \equiv   \( A e^{i\frac{3\pi\nu}{4}}   +  B e^{-i\frac{\pi\nu}{4}}\)   e^{i\zeta} ,
\label{a-}\\
\co_1^+ \equiv    i \sqrt{\nu} e^{i\theta_a} B \xi_{\nu-1}  e^{i\frac{5\pi\nu}{4}} e^{-i\zeta} ,\qquad
\co_2^+ \equiv    \(A e^{-i\frac{\pi\nu}{4}}   +  B e^{i\frac{3\pi\nu}{4}}\)   e^{i\zeta} ,
\label{a+}
\end{eqnarray}
where we used that
\be
	\frac{\Bc_{12}^{(0)}}{ \s N_1^{(0)}} \, \frac{|N_1^{(0)}|^{1/2}}{|N_2^{(0)}|^{1/2}} =\sqrt\nu e^{i \theta_a},\qquad \theta_a
		=\arg \frac{\Bcn0_{12}}{N_1^{(0)}}+\frac\pi4\(1-{\rm sgn}(N_1N_2)\),
	\label{theta_u}
\ee
which follows from  \Ref{sigma} and \Ref{nu_gen_s}.
The formulas \Ref{a-}, \Ref{a+} enable us to find the  transition matrix $\T$, as  a solution of the following equation valid for all values of $A$ and $B$
\be
 \T \(\begin{array}{c}
		 \co_1^-\\
		 \co_2^-
		\end{array}\)
=	\(\begin{array}{c}
		 \co_1^+\\
		 \co_2^+
		\end{array}\).
\label{trMa}
\ee
It reads
\be
\T
= \(\begin{array}{cc}
		 e^{ i\pi\nu} &
			\frac{i \sqrt{2\pi\nu} e^{ i\frac{\pi\nu}{2}} }{e^{2i\zeta-i\theta_a} \Gamma(1-\nu)} \\
		\frac{ \sqrt{2\pi\nu}\, e^{ i\frac{\pi\nu}{2}}e^{2i\zeta-i\theta_a}}{\Gamma(1+\nu)}&
			 e^{ i\pi\nu}
		\end{array}\),
\label{T}
\ee
here we used the identities
$\xi_{\nu-1}
=\frac{\sqrt{2\pi}}{\Gamma(1-\nu)} e^{-i\pi\nu}$ and
$\Gamma(1-\nu)\sin \pi \nu 
=\frac{\nu \pi}{ \Gamma(1+\nu) }$.

As it must, the obtained transition matrix  \Ref{T} satisfies general properties \Ref{gen-prop-T} discussed in Appendix \ref{app:T_prop} for any phase factor $e^{i\zeta}$. To check this fact, one should take into account that $\overline{\Gamma(1+\nu)} = \Gamma(1-\nu)$ and  \Ref{nu_gen_s}.

\subsection{Asymptotics of $\T$ and its canonical form}\label{sec:can-matr}

As we mentioned in Section \ref{adia_rearr}, we are free to choose the phase factor $e^{i\zeta}$ at will,  changing the definition of  adiabatic modes. For instance, we can search for such modes that
transform to each other with unit probability in the appropriate limit of degeneracy lifting. In other words, they must give the asymptotics of $\T$ for $|\nu| \gg 1$  in the simplest form.
This  enables us to fix uniquely the phase factor  $e^{i\zeta}$, thus
completing the unambiguous definition of the adiabatic modes \Ref{WKB1} (up to a constant factor common for all modes).

Before considering degeneracy lifting, we shall check the opposite limit of vanishing perturbation, $\nu\to0$. As expected in Section \ref{sec:numbering}, we have
\be
\T
\seq_{\nu\to0} \(\begin{array}{cc}
		 1\ &
			\ 0\\
		0\ &
			\ 1
		\end{array}\)
\label{T-unit}
\ee
for any choice of $\zeta$ and both for the avoided crossing case and real turning points.

To consider the $|\nu| \gg 1$ asymptotics, we start with Stirling's formula \cite{abramowitz}
\be
\Gamma(1+\nu) \seq \sqrt{2 \pi \nu} \( \frac{\nu}{e} \)^{\nu}, \quad  |\nu| \gg 1, \quad \arg{(\nu)} < \pi
\ee
for  $\nu = i |\nu|   \sgn (N_1   N_2)$, see \Ref{nu_gen}, and  obtain
\be\label{Gam-as}
 \Gamma(1+\nu) \seq_{|\nu| \to \infty} \sqrt{2 \pi|\nu|}\,
		 e^{-|\nu|\frac{\pi}{2}}
\, e^{i  \theta_{\Gamma}(\nu)}   , \quad
	i \theta_{\Gamma}(\nu) =  - \nu + \nu \ln|\nu| +i \frac{\pi}{4}{\rm sgn}({\small {  N_1   N_2}} ).
\ee
Note that  $\arg{\Gamma}(1-\nu) = - \arg{\Gamma}(1+\nu)$.  Now we may simplify the off-diagonal terms  $t_{12}$ and $t_{21}$  of $\T$ as follows:
\be
t_{12} \seq
-\frac{e^{-i\frac{\pi}{2}(1-{\rm sgn} (N_1 N_2)) (1+i|\nu|)  }}
	{ e^{ \nu - \nu \ln|\nu| +2i\zeta-i\theta_a }},  \qquad
t_{21} \seq
 \frac{ e^{- |\nu|\frac{\pi}{2}( {\rm sgn} (N_1 N_2)  - 1 )  }  }
	{e^{ - \nu + \nu \ln|\nu|   -2 i\zeta+i\theta_a}}.
\ee
By choosing the phase factor ${i\zeta}$ as
\be\label{zeta}
	i\zeta=\frac{i\theta_a}2 -\frac{\nu}{2} + \frac{\nu \ln|\nu|}{2},
\ee
we obtain the simplest possible form of the transition matrix asymptotics for $|\nu|\gg 1$
\be
\T \seq_{|\nu| \to \infty} \(\begin{array}{cc}
		 0\ & - 1 \\
		1\ & \phantom{-} 0				\end{array}\)  \quad{\rm if\ }    N_1 N_2 >0,  \qquad
\T \seq_{|\nu| \to \infty} e^{\pi |\nu|}\(\begin{array}{cc}
		   1\  & 1 \\
		 1\ &    1	
		\end{array}\) \quad {\rm if\ }     N_1   N_2 <0.
\label{T-lim}
\ee
The form of matrix $\T$ in the case of avoided crossing, $N_1   N_2 >0$,
reflects the peculiarities of the numbering method that we choose, see discussion in Section \ref{sec:numbering}. Note that the phase factor $-1$ cannot be changed at will by choosing the phase of adiabatic modes, since it would also change \Ref{T-unit}.

Actually, this phase is an intrinsic property of the transitions in the avoided crossing case. In the Appendix \ref{app:outer-lift}, we reveal the correspondence between the perturbed and unperturbed adiabatic modes in this limit,  which is given exactly by the first of the matrices \Ref{T-lim}.

Substituting \Ref{zeta} into \Ref{T} we get the final result
\be
\T = \(\begin{array}{cc}
		 e^{ i\pi\nu} &
			\frac{i \sqrt{2\pi\nu} e^{ i\frac{\pi\nu}{2} +\nu - \nu \ln|\nu| } }{\Gamma(1-\nu)} \\
		\frac{ \sqrt{2\pi\nu}\, e^{ i\frac{\pi\nu}{2} -\nu + \nu \ln|\nu|}}{\Gamma(1+\nu)}&
			 e^{ i\pi\nu}
		\end{array}\).
\label{final-T}
\ee
Note that $\sqrt{\nu} = e^{i \frac{\pi}{4} {\rm sgn}(N_1 N_2 )}\sqrt{|\nu|}$.

The adiabatic modes with $\zeta$ given by \Ref{zeta} we call \emph{canonical modes}, their explicit form is discussed in the following subsection. The transition matrix $\T$ \Ref{final-T}  between such canonical modes
we call  \emph{canonical transition matrix}.

\vskip3mm
Finally, we present the matrix $\T$ in another form, writing explicitly the absolute values and arguments of matrix elements. To this end, we take into account the fact  that $|\Gamma(1+\nu)|^2 = |\Gamma(1-\nu)|^2 = \pi |\nu|/\sinh{\pi |\nu|}$ and obtain
\be
\T = \(\begin{array}{cc}
		 e^{ -\pi|\nu|w} &
			 \sqrt{1 - e^{-2 \pi |\nu|}}  e^{ -\frac{\pi}{2} |\nu|(w -1)}  e^{ i\theta'}
e^{i \frac{\pi}{2}(1+w)}  \\
		 \sqrt{1 - e^{-2 \pi |\nu|}}  e^{ -\frac{\pi}{2} |\nu|(w -1)}  e^{ -i\theta'}&
			 e^{ -\pi|\nu|w}
		\end{array}\),
\label{final-sq-T}
\ee
where
$ w=\sgn N_1 N_2 $, and
\be\label{theta-1}
\theta' = \arg{\Gamma(1+\nu)} - \theta_{\Gamma},
\ee
and $\theta_{\Gamma}$ is the principal term of the asymptotics of $\arg{\Gamma(1+\nu)}$  if $|\nu| \to \infty$, see \Ref{Gam-as}, therefore $\theta' \to 0$.

We note that the obtained matrix $\T$ \Ref{final-T} describes the transition process between the principal approximations of the adiabatic expansion \Ref{WKB1}. Thus, it can be called the principal approximation of the complete (exact) transition matrix of the problem.


\subsection{ {Canonical adiabatic modes}\label{sec:can-modes}}
We have found above the transition matrix \Ref{final-T} between two adiabatic modes \Ref{WKB1}, whose remaining ambiguity in the overall phase factor was fixed by \Ref{zeta}. We called such modes the canonical ones.

Substituting  \Ref{zeta} into \Ref{Theta1} and then  into  \Ref{WKB1}, we obtain their final form, which however contains a number of diverse terms in the phase factor without a clear physical meaning. Recalling an observation made after \Ref{rearr_ph}, we can interpret them as an integral of the perturbed eigenvalue $\hb$ \Ref{eig-root} in the vicinity of the degeneracy point.

Indeed, by explicit integration and trivial series expansion at the upper limit of integration, we find
that for $\tau^*=x^*/\sh\sim\h^{-\g}$ with $\g\in(0,1/2)$
\be
i \int^{\tau^*}_{\t_{\pm}} d\tau' \sqrt {Q^2 (\tau' + b)^2	- 2 i \nu Q}
	=
  \frac{i Q (\t^* +b)^2}{2}  + \nu \ln|\s \tau^*|  + \frac{\nu}{2} - \frac{\nu \ln|\nu|}{2} +O(\h^{1/2-\g})
\label{tau*int}
\ee	
holds.
The lower limit of integration is chosen to be equal to the real part of the degeneracy point \Ref{deg-point}
\be
\t_{\pm}\equiv \Re{\varkappa_{\pm}}/\sh , \quad	\t_\pm = \left\{\begin{array}{lll}
 - b &
		& {\rm if\ } N_1 N_2 >0, \\
 -    b \pm    p/ Q
		&  &{\rm if\ }  N_1 N_2 <0.
\end{array}\right.
\label{tau pm}
\ee

Combining now \Ref{WKB1}, \Ref{Theta1}, and \Ref{zeta} with \Ref{tau*int}, we obtain
\be
	i \Theta_1=  \frac{i}\sh\int^{\t^*}_{{-b}} \(\b_0 + \sh \hb_{av}^{(1)}(\t')\) d\t'
	- i \frac{Q (\t^* +b)^2}{2}  -  \nu \ln|\s \tau^*| + \frac{i\theta_a}2 -\frac{\nu}{2} + \frac{\nu \ln|\nu|}{2}.
\label{Theta1-can}
\ee
 In \Ref{tau*int}, we immediately recognize the  terms outside the integral from \Ref{Theta1-can},  and, introducing the notation $\hb_j^{pr}(x',\h)$ and $\vartheta_j$, we get
the following form for the canonical mode:
\be
 {\Ps}_{j\pm}^{(0)} =  e^{i\vartheta_j}\,
  \frac{{  \vph_j}(x)}{\left|( \vph_j(x),\G  \vph_j(x))\right|^{1/2} }
  \exp{\frac{i}\h \,\int\limits^x_{\Re{\varkappa_{\pm}}} \,\(  \hb_j^{pr}(x',\h) -  \h \Im S_{jj}(x')\) dx' },
 	\label{Psi-can}
\ee
where
 \be
  \hb_j^{pr}(x',\h) = \left\{
  \begin{array}{ll}
 \b_j+\sqrt\h \frac{ \Bc_{jj}}{  N_j}
 +\h\[\frac{ (-1)^j \Bc_{\brr21}  \Bc_{\brr12}}{ ( \b_2- \b_1)   N_1  N_2 }
	+\frac{  ( \vph_j,\B  \vph_{j\perp}^{(1)}) }{  N_j} \]
		& {\rm if\ } |x'|\in (x^*, |x| ), \\
 \b_0 + \sh \b_{av}^{(1)} +  (-1)^j\,\h \,{\rm sgn}(\frac{x'}{\sh} + b) \sqrt {Q^2 (\frac{x'}{\sh} + b)^2	- 2 i \nu Q}\quad
		  &{\rm if\ }  x'\in (-x^*,x^*), \label{hbet-pr}
\end{array}
\right.
  \ee
with arbitrary $x^* \sim\h^{1/2-\g}$ with $\g\in(0,1/2)$ and
\be\label{Thet-chos}
\vartheta_j=
		 (-1)^{j+1} \frac{\theta_a}2
		 +\frac{1}{\sh} \int\limits_{{-b}}^{\t_{\pm}} \(\b_0 + \sh \hb_{av}^{(1)}(\t')\) d\t'.
\ee
Here $\t_{\pm}$ and $\theta_a$ are determined by \Ref{tau pm} and \Ref{theta_u}, and $\b_0$ and $\hb_{av}^{(1)}$ are defined in \Ref{b-av} and \Ref{b-0-def}, respectively. An integral term in the formula (\ref{Thet-chos}) vanishs if $N_1 N_2>0$.

The function $\hb_j^{pr}(x',\h)$ can be regarded as the first terms of the asymptotic expansion of the perturbed eigenvalue $\hb_j$ as $\h \to 0$, depending on the considered interval of the values of $x'$: \Ref{hb} for $x'$ well separated from the degeneracy point, $x'\gg \sh$ (upper line of \Ref{hbet-pr})  and \Ref{eig-root} for smaller values of $x'$ (lower line of \Ref{hbet-pr}).
Note that $\hb_1^{pr}(x',\h) $ and its $x'$-derivative have jumps of order $O(\h^{1/2+\g})$ in $x'=x^*$, which are negligible in the $O(\h^{1/2})$ approximation.
Moreover, $\hb_1^{pr}(x',\h) $ is independent of $x^*$ in the same approximation.
 We recall that it is the transition process between these modes that is described by the canonical $\T$ given in \Ref{final-T}.

Finally we note that by selecting phases of $\vph_j (0),$ $j=1,2$, we can always obtain $\theta_a=0$. Indeed,  if we replace $\vph_1 (0)$ and $\vph_2 (0)$ by $e^{\frac{i\theta_a}2}\vph_1 (0)$ and $e^{\frac{-i\theta_a}2}\vph_2 (0)$, respectively, ${\Bc_{12}^{(0)}}$ is replaced by a real positive quantity for $N_1 N_2 >0$, and a purely imaginary one if $N_1 N_2 <0$. In both cases, $\theta_a$ vanish as follows from \Ref{theta_u}. The same relation can be obtained by imposing a condition on the matrix elements of $\B$:
\be\label{phase-condition}
	\frac{\Bc_{12}}{N_1}
	=\frac{\Bc_{21}}{N_2}.
\ee


\subsection{Transition matrix for arbitrary modes}\label{sec:arb-modes}

Consider the adiabatic mode of the general form \Ref{WKB1}, $j=1,2$:
\be
 \wt{\Ps}_{j\pm}^{(0)} =
  {  \vph_j}\,\,
e^{\frac{i}\h \,\int\limits^x_{x^{\pm}_{j}} \,
	\(
 	 \b_j
 	+\sqrt\h \frac{ \Bc_{jj}}{  N_j}
	 +\h\[
	 	\frac{ (-1)^{j }\Bc_{\brr21}  \Bc_{\brr12}}{ ( \b_2- \b_1)   N_1  N_2 }
		+\frac{( \vph_j,\B  \vph_{j\perp}^{(1)})}{  N_j}  + i S_{jj}
		\]
	\)dx'}
\label{Psi12a1}\ee
$$=
|N(x^{\pm}_{j})|^{1/2}
  	\frac{{  \vph_j}(x)}{|N_j(x)|^{1/2} }
	e^{\frac{i}\h \int\limits^x_{x^{\pm}_{j}}
	\(
 	 \b_j
 	+\sqrt\h \frac{ \Bc_{jj}}{  N_j}
	 +\h\[
	 	\frac{ (-1)^{j }\Bc_{\brr21}  \Bc_{\brr12}}{ ( \b_2- \b_1)   N_1  N_2 }
		+\frac{( \vph_j,\B  \vph_{j\perp}^{(1)}) }{  N_j} -  \Im S_{jj}
		\]
	\) dx' },
$$
with an appropriate lower limit of integration lying in the area of applicability of the solution on the same side of the degeneracy point $x=0$, $\sgn x_j^\pm = \sgn x$, $|x_j^\pm| >> \sh$.
The eigenfunctions ${ \vph_j}$ have arbitrary normalization and phase factor.

Comparing  \Ref{Psi12a1} with \Ref{Psi-can}, we write
 \be\label{def-Ps-tld}
	\wt\Ps^{(0)}_{j \pm}=n_{j}^\pm \Ps^{(0)}_{j \pm}, \quad j=1,2,
\ee
where the constant factors are
  \be
 n_j^{\pm} = |N_j(x^{\pm}_{j})|^{1/2}\, e^{-i\vartheta_j} \,
 	\exp{  \frac{i }\h\,\int\limits_{x^{\pm}_{j}}^{\Re{\varkappa_{\pm}}}
 		\(\hb_j^{pr}(x',\h)
 		-   \h\Im S_{jj}(x') \) dx' }, \label{n_j}
 \ee
where the notation is from \Ref{hbet-pr} and \Ref{Thet-chos}.

The new adiabatic coefficient $\tilde \co$ in \Ref{ps-infty-}, \Ref{ps-infty+} will also be altered in the following way
\be
	\wt{\co}_j^- = \frac{\co_j^-}{n_j^- }, \qquad
	\wt{\co}_j^+ = \frac{\co_j^+}{n_j^+}, \qquad j=1,2.
\ee
As compared to \Ref{T} the new transition matrix will read in this case
\be\label{trans-bas}
\wt{\T}
 = \(\begin{array}{cc}
		   {n_1^+} & 0 \\
			0 & {n_2^+}
		\end{array}\) \T
	\(\begin{array}{cc}
		  \frac1{n_1^-} & 0 \\
			0 &   \frac1{n_2^-}
		\end{array}\).
\ee
This is the main formula for expressing the transition matrix   through the canonical one  in the general case.

\section{Physical interpretation}\label{sec interp}

The transition matrix written above enables one to reconstruct formally the adiabatic solution on a given side of the degeneracy point, provided that we know it on the other side. Now we proceed with the physical interpretation of  results obtained. We recall  that for any solution $\Ps$, the quantity $(\Ps , \G\Ps)$ is a constant, see \Ref{conser}; it does not depend on $x$ and for problems of waves propagation (see Appendix \ref{app:examples}) has the meaning of the time-averaged flux of energy.

For physical interpretation of the transition process described by matrix \Ref{trans-bas}, we must first establish the direction of modes propagation, and thus identify the incident and reflected waves.
One of the possibilities to determine it is by the sign of $N_j, j=1,2$, i.e., by the direction of the time-averaged flux of energy. Another way to do it is based on the sign of the  phase speed, which most useful when $\sgn N_1=\sgn N_2$. In the problems originated in  monochromatic waves propagation, the harmonic factor, $e^ {-i\frac{\omega t}{\h}}$, is usually omitted and the velocity of phase propagation is $v_{j} = \omega/\hb_j$. Therefore for such problems, the sign of the phase velocity coincides with the $\sgn \hb_j$, while $\sgn N_j$ coincides with the one of the group velocity.
  The direction of the phase velocity and the energy flux may not coincide.

Now let us consider  two adiabatic solutions $\Ps_j$, $j=1,2$, with  asymptotics  \Ref{WKB1}, say, for negative $x$ corresponding to real eigenvalues $\b_1$ and $\b_2$ degenerating at $x=0$. Depending on the signs of $(\Ps_1 , \G\Ps_1)$ and $(\Ps_2 , \G\Ps_2)$, the transition of these solutions would correspond to two different physical processes.

First, let the signs of $N_j, j=1,2$,  be opposite. This means that solutions $\Ps_j$, $j=1,2$ describe fields with opposite direction of energy fluxes.  As shown in \Ref{eig-root}, in this case the perturbed problem has two real degeneracy points.

The sign of $N_j$ may be both positive and negative for a mode of any number. Let $N_1>0$, $N_2<0$.
Then the first mode for negative $x$ may be regarded as an incident one and the second mode as the reflected one from the degeneracy points.  The first mode behind the turning points for positive $x$ is a transmitted one.
Putting in the definition of the transition matrix \Ref{trMa} $\co_1^-=1$, $\co_2^-=R$, $\co_1^+=T$, $\co_2^+=0$, we obtain for the reflected $R$ and transmitted $T$ coefficients the following:
\be
	R=-\frac{\Tc_{21}}{\Tc_{22}},\qquad
	T=\frac{\det \T}{\Tc_{22}}.
\label{rt}
\ee
If the modes are canonical  \Ref{Psi-can}, the transition matrix
 \Ref{final-sq-T}  for $N_1 N_2=-1$  gives
\be
  R =- \sqrt{1 - e^{-2 \pi |\nu|}} e^{ -i\theta'}, \quad
  T = e^{- \pi |\nu|},
\ee
where $\theta'$ is defined in \Ref{theta-1}. If the modes are not canonical, the matrix elements of $\wt\T$ \Ref{trans-bas} should be used in \Ref{rt} instead of those of $\T$.

If $N_1$ and $N_2$ have the same sign, then the direction of energy propagation, or its analog,  for both degenerating modes is the same.
In this case, we have the avoided crossing of eigenvalues of the perturbed problem as  may be seen from
\Ref{eig-root}, and we interpret the process near  $x=0$ as the transformation or interaction of adiabatic modes.
The first mode incident from the left, i.e., $\co_1^-=1$, $\co_2^-=0$ in \Ref{ps-infty-}, produces two modes on the right of $x=0$ with  $\co_1^-=t_{11}$, $\co_2^-=t_{21}$, where the corresponding elements of the matrix $\T$ are the transmission coefficient $t_{11}$ of the first mode,  and the excitation coefficient $t_{21}$  of the second one. The matrix  \Ref{final-sq-T} yields
 \be
\T = \(\begin{array}{cc}
		 e^{ -\pi|\nu|} &
			- \sqrt{1 - e^{-2 \pi |\nu|}}    e^{ i\theta'} \\
		 \sqrt{1 - e^{-2 \pi |\nu|}}   e^{ -i\theta'}&
			 e^{ -\pi|\nu|}
		\end{array}\).
\label{final-T-LZ}
\ee

It is worth noting that the numbering of modes introduced by us  is not convenient in the case of weak interaction of modes, i.e., in the case where perturbed eigenvalues are separated so that the adiabatics may be applied. In our numbering, the eigenvalues of the adiabatic modes have an abrupt jump, see thick lines in Fig.~\ref{compl_TP}, left.
If we change the numbering of modes on the left of the degeneracy point, i.e., if we identify the incident and reflected waves by their phase velocities, then the eigenvalues for large $|\nu|$ will be continuous and the transition matrix for the new numeration reads
\be
\T \(\begin{array}{cc}
					 0  &  1 \\
			         -1  &  0		\end{array}\)
= \(\begin{array}{cc}
					 \sqrt{1 - e^{-2 \pi |\nu|}}    e^{ i\theta'}  &  e^{ -\pi|\nu|} \\
			 -e^{ -\pi|\nu|} &   \sqrt{1 - e^{-2 \pi |\nu|}}   e^{-i\theta'}
		\end{array}\).
\label{number-T-LZ}
\ee



\section{Conclusions }\label{sec:concl}
In the present paper, we studied a special class of Schroedinger equation with non-Hermitian Hamiltonian, permitting a representation \Ref{factor}. We showed that such a formulation incorporates a wide range of very different physical problems,
arising in quantum mechanics or in problems of waves propagation (such as elastic waves, radiowaves, and the others).

The special form of the Hamiltonian yields a conservation law: the quadratic form  $(\Ps , \G\Ps)$ is constant for any solution $\Ps$. The sign of this form is an intrinsic property of solutions. For problems of waves propagation, which we considered, it has the meaning of a flux of energy.
The relative sign of the fluxes of degenerating modes, $w\equiv\sgn (\Ps_1 , \G\Ps_1)(\Ps_2 , \G\Ps_2)$, determines the character of the transition --- either it follows the avoided crossing scenario ($w<0$), or the real turning points one ($w>0$). Both regimes were considered within a unified asymptotic approach.

We constructed formal adiabatic expansions  for the equation \Ref{mainEq0} for $\h \to 0$, assuming that two modes of the unperturbed operator degenerate at $x=0$. The perturbation causes the splitting of the point of  degeneration  (degeneracy point)  into two real or complex simple degeneracy (turning) points dependent on $w$.

Employing the method of matched asymptotic expansions, we solved the transition  problem  by passing through  a neighborhood of the degeneracy point,  constructing an inner solution, which incorporates the contributions of both degenerating modes. It contains  parabolic cylinder functions, and  the transition matrix was obtained by matching their asymptotics to adiabatic modes.

Generally speaking, the transition matrix depends on the normalization of adiabatic modes and their relative phases.
Fixing all these ambiguities, we obtained a canonical transition matrix, having the most natural asymptotics for vanishing perturbation and in the limit of lifting the degeneracy. It depends on the relative sign of the fluxes $w$ and on the parameter $\nu$, see \Ref{nu-rep}, which is  determined by the eigenvalues behavior and can be found numerically if needed.

The transition matrix depends only on eigenvalues and eigenfunctions of two degenerating modes and does not depend on the rest of the spectrum, as is prescribed by the adiabatic theorem.  If $w$ is fixed,
the absolute values of the canonical matrix entries, which are responsible for energy relations, are completely determined by eigenvalues behavior  of the perturbed operator. The  results obtained coincide with those of Landau and Zener \cite{landau_32}, \cite{zener_32} and \cite{LZ_hagedorn_91} for the avoided crossing scenario and \cite{koen}, \cite{zalip} for the two real turning points scenario.   The only restriction on the choice of modes to obtain the absolute values of the matrix entries is that they  should have the same normalization.

If, in addition, the eigenfunctions of the unperturbed operator  are known at the degeneracy point, the phases of adiabatic modes can be fixed by \Ref{Thet-chos} and the phases of the transition matrix entries  can be obtained. They coincide with those found in \cite{bobashov}, \cite{bobashov2} for the case of an avoided crossing and with \cite{zalip} for two real turning points.

The usual way of treating the non-Hermitian problems consist of dealing with a non-Hermitian Hamiltonian $\H$ and constructing biorthogonal bases  $\{\vph\}$, $\{\vph^\dag\}$ of eigefunctions of $\H$  and $\H^\dag$ correspondingly. In such approaches, the normalization of a solution $(\Ps^\dag, \Ps)$ can always be chosen positive, since there is no relation between  $\vph$ and $\vph^\dag$ presupposed, and the latter one can always be supplied with an appropriate phase factor. In this case, the nature of  perturbed degeneracy points (either avoided or real crossing of eigenvalues will be invoked) is fully defined by the (non-Hermitian) perturbation operator $\B$, and its matrix elements $(\vph^\dag, \B\vph)$.

The connection of our approach with the aforementioned one follows from the relation $\vph^\dag= \a \G\vph$, which is valid for the class of  Hamiltonians \Ref{factor}. We put $\a=1$, thus fixing the ambiguity in the  definition of the adjoint eigenfunction. The advantage of our approach is that the nature of the perturbed degeneracy is revealed by simple comparison of the signs of the normalization of the degenerating modes, $\sgn (\Ps_i, \G\Ps_i)=\sgn (\vph_i, \G\vph_i)$, $i=1,2$.

We believe that the suggested  approach enables to get the results in a physically more transparent manner. Our considerations can be generalized to the case where $\G^{-1}$ does not exist or $\G$ is not a matrix but, for example, a differential operator.
Further generalizations of the approach to the multidimensional case, i.e., to the equation of the form of
\be
	\(\K(x,y)+\delta \B(x,y)\)\Ps= -i\h \left( \G_x\frac{\partial \Ps(x)}{\partial x}
+ \G_y\frac{\partial \Ps}{\partial y} \right)
	\label{main-multi}
\ee
ares also possible, and are the subject of the future work. An example of treatment of
\Ref{main-multi} is given in  \cite{PerelSidorenko-crystal}.

The extraction of $\G$ can also reflect particular symmetries of a  given  physical problem, as is for the Dirac equation, see \Ref{dirac}. Another example of the applications of the equation in the form \Ref{mainEq0}
is the case of ${\cal PT}$-symmetric quantum mechanics with $\G$ playing the role of a ${\cal P}$-symmetry operator (see \cite{Bender2007}).
We shall note that some aspects of the presence of degeneracy points, also called \emph{exceptional} ones \cite{BerryPT04,Heiss12},  were considered, e.g.,  in \cite{Andrianov07,Reyes12}. However, only the case of  complex degeneracy points (i.e., avoided crossing) were dealt with, since the presence of real degeneracy points would signal entering the broken ${\cal P}$-symmetry region.
Considering the transition of a quantum system through such a region is yet another application of the general method presented here.

\section*{Recipe}
\label{sec:recipe}

We conclude this paper with a recipe on solving the connection problem without necessity of following all the intermediate steps presented in the preceding sections.

The general procedure is the following. For a given physical problem where a degeneracy point is present of the considered type, one should first define unambiguously the adiabatic modes, and then calculate the parameter $\nu$, which defines the transition matrix $\T$.
This can be carried out by performing the following steps.
\begin{itemize}
\item First of all, one should rewrite the physical problem in hand in the form
\be
	(\K+\sqrt\h \B) \Ps=-i\h\partial_x\G \Ps
\label{H+B}
\ee
where $\K$, $\B$, and $\G$ are selfadjoint operators not containing the derivatives in $x$, and $\h$ is a parameter that can be considered small in the physical problem at hand.

Despite there is no universal way how to accomplish this step for a given system of linear partial differential equations, in many cases the approaches used in \cite{felsen_marcuvitz_72,ion_perel_90,timo_perel_00} and Appendix \ref{app:examples} yield the required result. Usually, $\K$ and $ \B$ will be given by square  matrices with elements containing derivatives in coordinates other than $x$, while first component(s) of the vector $\Ps$ will be given by original unknown function(s), and the others will be given by linear combination of its (their) derivatives.
We note that the choice of  linear combinations of  derivatives of unknown functions as the components of $\Ps$ may be suggested by  considering in detail the conservation law inherent for the physical problem --- it must transform  into the bilinear form $(\Ps, \G \Ps)=const.$
\item
At the next stage, one should solve the generalized eigenvalue problem for $\K(x)$:
\be
	  \K (x )  \vph_j(x) = \b_j(x) \G  \vph_j(x), \quad j=1,2,
	  \label{newEVP1a}
\ee
and check that its solutions indeed satisfy the conditions assumed in this paper: there is only one  degeneracy point of the eigenvalues
\be
	 \b_2(x) -  \b_1(x) \seq_{x\to0} 2 Q x,\qquad
	Q>0,
	\label{11}
\ee
and the corresponding eigenfunctions $\vph_j$ are linear independent at this point.  We note that $\vph_j(0)$ should be defined by continuity at $x=0$.
The ``normalization" constants of eingenfunctions
\be
	  N_j=(\brr{ \vph_j},\G \vph_j), \qquad
	j=1,2,
\label{Ni}
\ee
should be calculated. The eigenfunctions may be always normalized in such a way that $|N_j|=1$  at the degeneracy point, but we do not assume it was accomplished. The sign of $N_1 N_2$ is an intrinsic property of the problem and cannot be changed by any alteration of the eigenfunctions: if $N_j, j=1,2$, are of the same sign, the system follows the avoided crossing scenario, Fig.~\ref{compl_TP}; otherwise, it follows the real turning points one, Fig.~\ref{real_TP}.

\item
Next, one must to construct the (principal term of) adiabatic expansion in the outer regions, $\wt{\Ps}_{j\pm}$ on the right/left of the degeneracy point,%
\begin{eqnarray}
&&  \wt{\Ps}_{j\pm} =
  {  \vph_j}\,\,
  e^{i \,\int\nolimits^x_{x^\pm_{j}} \,\( \h^{-1} \hb_j^{pr}(x',\h) +i   S_{jj}(x')\) dx' }
, 	\label{Psi12a}\\
&& \hat{\b}_j^{pr}(x',\h) =  \b_j+\sqrt\h \frac{ \Bc_{jj}}{  N_j}
 +\h \[ \frac{  \Bc_{\brr21}  \Bc_{\brr12}}{ ( \b_1- \b_2)   N_1  N_2 }
	+\frac{( \vph_j,\B  \vph_{j\perp}^{(1)}) }{  N_j} \], \label{b-as}
\end{eqnarray}
with any lower limit of integration, which is appropriate for the physical problem, but always lying on the same side of the degeneracy point $x=0$ as the area of applicability of the corresponding solution , $x^\pm_{j}\sim x$, $|x|\gg \sh$.
Here $\hat{\b}^{pr}_1(x',\h)$ is an eigenvalue of $\K+\sqrt\h \B$ taken  with sufficient precision for calculation of the principal term of the asymptotics. Instead of using analytical expression \Ref{b-as}, one can also find it numerically.

\item
The (nonadiabatic) transformation of adiabatic modes \Ref{Psi12a} in a neighborhood of a degeneracy point is solved then by the matrix $\tilde{\T}$ in the following sense. By a given linear combination of modes \Ref{Psi12a} on one side of the degeneracy point (say, left),
\be
 	\Ps
	 \mathop{\simeq}_{{x \ll -\sh}}
		 \tilde{\co}_1^- \wt{\Ps}^{(0)}_{1- } +    \tilde{\co}_2^-\wt{\Ps}^{(0)}_{2- },
\ee
we find a linear combination on the other side (say, right),
\be
	\Ps
	 \mathop{\simeq}_{{x \gg \sh }}
		 \tilde{\co}_1^+ \wt{\Ps}^{(0)}_{1+ } +    \tilde{\co}_2^+ \wt{\Ps}^{(0)}_{2+ }, \label{R-ps-infty}
\ee
by using the following equation:
\be
\(\begin{array}{c}
		 \tilde{\co}_1^+\\
		 \tilde{\co}_2^+
		\end{array}\)=
\wt{\T} \(\begin{array}{c}
		 \tilde{\co}_1^-\\
		 \tilde{\co}_2^-
		\end{array}\)
\label{T_gen}
\ee
with $\wt\T$ given by formula \Ref{trans-bas}
  with account of  \Ref{final-T}, \Ref{Thet-chos} and \Ref{n_j}.

The transition matrix is fully determined by a single parameter $\nu$,
\be\label{nu-rep}
	\nu = i  \frac{p^2 {\rm sgn}(N_1 N_2)}{2Q},
\ee
which is a dimensionless ratio of two physical parameters of the problem: the separation of perturbed eigenvalues near the degeneracy point or distance between turning points, $p$, \Ref{not-p}, and the eigenvalue's relative inclination $Q$, see \Ref{11}.

Upon completing the above steps, we obtain a transition matrix together with an  unambiguous definition of the adiabatic modes, which solves completely the physical problem.

\end{itemize}


\section*{Acknowledgements}
{This work was supported in part by FAPESP (I.V.F) and grants RFBR
140200624, 140200198, and  SPbGU grant 11.38.263.2014 (M.V.P.). }


\appendix




\section{Properties of the $\G$-eigenvalue problem} \label {app:eig-pr}
The eigenvalue problem \Ref{sEVP}
\be
	\K\vph=\b\G\vph
	\label{sEVPa}
\ee
with $\G \ne \I $ ($\I$ is the identity matrix) has different properties, as compared with the case of $\G =\I$.

\begin{enumerate}
\item{
First of all, the eigenvalues may be both real and complex. The complex eigenvalues appear in pairs: $\b_n$ and its complex conjugate $\overline{\b_n}\equiv \b_{\n}$. The index $\n$ means the sequential number of the eigenvalue $\overline{\b_n}$.  The eigenfunctions corresponding to $\overline{\beta_n}$ are denoted by $\vph_{\n}$.}
\item{
 The normalization factor for a real eigenvalue $\b_n$
\be\label{norm-def}
N_n = (\vph_n, \G \vph_n)
\ee
is real, but may be positive and negative.
This causes new properties of solutions. We note that it is always positive
if $\G= \I$ and $\vph_n \ne 0$.

We note that for complex eigenvalues $\b_m$ the normalization factor is $(\vph_{\overline{m}}, \G \vph_m)$.}

 \item{ If the eigenvalue $\b_n$ is not degenerate, we have
 \be
 N_n \ne 0.
 \ee
The same is true if $\b_n$ is twofold degenerate, and the algebraic multiplicity is equal to  the geometric one.

 }
 \item{
 The eigenfunctions of the numbers $m$ and $n$ satisfy the orthogonality conditions
\be\label{ort}
(\vph_n,\G \vph_m) =0 , \quad
{n} \ne m, \overline{m}.
\ee
}
\item{
Let $\b_1$ and $\b_2$ be  eigenvalues of the problem \Ref{sEVPa}
separated from  the other spectrum with a gap. The corresponding eigenfunctions are $\vph_j$, $j=1,2$.

 If
$(\vph_{j},\G\f)=0$, $j=1,2$, the
equation
\be\label{cor-lm2}
	(\K-\b_1\G)\vph_{1\perp} =
		 \G \f
\ee
has a unique solution $\vph_{1\perp}$ such that  $(\vph_{j},\G\vph_{1\perp})=0$, $j=1,2$.
}
\end{enumerate}

The properties 1,2,4 are obtained in the standard way.
Taking the inner product of the equation (\ref{sEVPa}) for $\vph_n$ and the eigenfunction $\vph_n$, we derive
that $\b_n$ is real if $N_n \ne 0$, using the fact that  $(\vph_n,\K\vph_n)$ and $N_n$ are real. Taking the inner product
of the equation (\ref{sEVPa}) for $\vph_m$ and the eigenfunction $\vph_n$, we obtain
 \be\label{help-ort}
(\overline{\b_n}-\b_m) (\vph_n,\G \vph_m)=0.
 \ee
We used the fact that  $\K$ is selfadjoint and the property of the inner product
$(\b_n\vph_n,\G \vph_m)=\overline{\b_n} (\vph_n,\G \vph_m)$. The relation  (\ref{help-ort}) yields (\ref{ort}).

Now we proceed with a proof of the property 3, by using some general facts from \cite{Kato} formulated for our case.
\begin{itemize}
\item
For our operator $\G^{-1}\K$, where $\K$ and $\G$ are selfadjoint, $\G$ and $\G^{-1}$ are bounded, we can construct a projection ${\P}$ to the invariant subspace $M'=\P \cal{H}$, corresponding to an isolate part of the spectrum $\Sigma'$. In our case, $\Sigma'$ consists of a  single eigenvalue or a pair of degenerating eigenvalues separated from  the other spectrum with a gap. All the Hilbert space $\cal{H}$ is split into the direct sum
\be
{\cal H} = M' \bigoplus M'',
\ee
where $M''=(\I-\P)\cal{H} $. Then every vector $\f$ is represented as ${\f} = {\P}{\f}  + (\I-{\P}) {\f}$,
${\P}{\f} \in M'$, $(\I-\P) {\f} \in M''$.  Both subspaces $M'$, $M''$ are invariant  for the operator $\G^{-1}\K$.

\item
The projection $\P$ can be found as
an integral of the resolvent of  $\G^{-1}\K$ along the contour $C$, which surrounds an isolated part of its spectrum $\Sigma'$:
\be
{\P} = \frac{1}{2 \pi i} \int_{C} (\G^{-1}\K -\b {\I})^{-1} \, d\b.
\ee
The adjoint projection ${\P}^+$  is an integral of the resolvent of the adjoint operator along the contour $C'$, which surrounds an isolated part of the spectrum of the adjoint operator:
\be
{\P}^+ = \frac{1}{2 \pi i} \int_{C'} ((\G^{-1}\K)^+ -\b {\I})^{-1} \, d\b.
\ee

\end{itemize}

We note that if $\K$ and $\G$ are selfadjoint
\begin{enumerate}
\item
\be
(\G^{-1}\K)^+ = \K \G^{-1}
\ee
\item
The spectrum of the adjoint operator $\K \G^{-1}$ coincides with the spectrum of the operator $\G^{-1}\K$ itself; therefore the contours coincide, $C'=C$. The adjoint eigenfunctions $\vph_j^+$ are determined as follows: $\vph_j^+ = \G \vph_j. $

\item
The projections satisfy the relations
\be\label{adjoint-op}
{\P}^+ = \G {\P} \G^{-1},
\ee
 which follows from the relation  $ \K\G^{-1} - \b \I = \G (\G^{-1}\K - \b {\I}) \G^{-1}.$

\end{enumerate}

\vskip 24pt

 {\it Lemma 1}

 The subspaces $M'$ and $M''$ are $\G$ - orthogonal, i.e.,
 \be
 ({\f},\G {\g}) = 0 \quad{\rm if\ }  {\f}\in M',  {\g}\in M'',
 \ee

 \vskip12pt
  which follows from the relations
  \begin{eqnarray}
 && {\f} = {\P}{\f}, \quad {\g} = ({\I}-{\P}){\f},\\
 && ({\P} \f, \G(\I-\P)\f) = (\f, \P^+ \G(\I-\P)\f),\\
 && \P^+ \G(\I-\P)=\G \P \G^{-1}\G(\I-\P)= \G \P(\I-\P)=0.
  \end{eqnarray}

  \vskip 24pt
  {\it Lemma 2}

  If
  \be
 ({\f},\G {\q}) = 0
  \ee
 for all ${\f}\in M'$, then ${\q}\in M''.$

  \vskip 12pt

  Assume that $\q=\q'+\q''$, ${\q'}\in M',$ ${\q''}\in M''.$
Then
\be \label{f-G-q'}
({\f},\G {\q'}) = 0,
\ee
 because $({\f},\G {\q''}) = 0$ according to Lemma 1.
 A vector $\G \q'$ as  any vector from the Hilbert space can be represented as a sum $\G \q' = \f_1 + \g_1$, ${\f_1}\in M'$,  ${\g}_1\in M''.$
For $|\G \q'|^2$ we obtain  $|\G \q'|^2 = (\q',\G \f_1) + (\q',\G\g_1) = 0$ because of \Ref{f-G-q'} and Lemma 1.  We derive that $\q'=0$ and $\q=\q'' \in M''$.

  \vskip 24pt

  From  Lemma 2 we obtain the property 3.
  Indeed, let $(\vph_j,\G\vph_1)=0$ for $j=1,2$. Let the basis of $M'$ consists of  $\vph_1$ and $\vph_2$. Then $\vph_1$ belongs to $M''$, which  contradicts  the assumption that $\vph_1 \in M'$.

 Now consider a one-dimensional $M'$, which is $c \vph_1$ for any constant $c$. The subspace $M'$ has a single basis vector $\vph_1.$  If $(\vph_1,\G\vph_1)=0$, then $\vph_1$  belongs to $M''$ as well. We have a contradiction, because it belongs to $M'$.

\vskip 24pt
The property 5 follows from Lemma 2 as well. It is assumed that  $\f$ is $\G$ -- orthogonal to $M'$. Then $\f \in M''$ by Lemma 2. The subspace $M''$ is an invariant subspace for the operator $(\G^{-1}\K-\b_1 \I)$. The solution can be found in the form
$\vph_{1\perp}^{(n)} = (\G^{-1}\K-\b_1 \I)^{-1}_{\perp} \f$, where $(\G^{-1}\K-\b_1 \I)^{-1}_{\perp}$ is the restriction of the resolvent to $M''$.

\vskip 24pt
If $\K=\K(x)$, then the eigenvalues and eigenfunctions, generally speaking, also depend on $x$. Useful properties for such dependencies  and real eigenvalues  are
\begin{enumerate}
\item{
\be\label{db-dx-h}
 \frac{d \b_k(x)}{ d x} = \frac{\Kc'_{kk}(x)}{N_k}
\ee
}
\item{
\be\label{conver-matr}
 (  \b_k(x) -   \b_j(x))    S_{jk}(x) =   \Kc'_{jk}(x),
\ee}
\end{enumerate}
where
\be\label{H1-matr}
  \K'(x) = \frac{d  \K(x)}{d x}, \quad   \Kc'_{jk}(x) \equiv (  \vph_j(x),  \K'(x)   \vph_k(x)), \quad j=1,2,\quad k=1,2,
\ee
and the conversion coefficients $S_{jk}(x)$ are defined in \Ref{def-B-perp}.

Indeed, differentiating the eigenvalue equation (\ref{newEVP}) for the $ \vph_k$ with  respect to $x$ and scalar multiplying it by $  \vph_j$, we get
\be\label{cond-help}
	( \vph_j(x),  \K'(x)  \vph_k(x)) = \(  \b_k(x)-\overline{ \b_j(x)}\) (  \vph_j(x), \G \frac {d \vph_k(x)}{d x})
+ \frac{ d \b_k(x)}{d x}   N_k \,\delta_{jk},
\ee
where $\delta_{jk}$ is the Kronecker symbol.
If $j=k$, we obtain \Ref{db-dx-h}.  If  $j\ne k$ and the eigenvalues are real, we have \Ref{conver-matr}.

Formulas \Ref{db-dx-h} and \Ref{conver-matr} have a useful consequence.
The matrix entries of $\K'$ satisfy the condition
\be\label{H12}
  \Kc'_{12} (0)= \Kc'_{21} (0) = 0.
\ee
 at the degeneracy point $x=0$.

This follows from (\ref{conver-matr}) if we take into account the fact that
the conversion coefficient  $  S_{jk}(x)$ is bounded for any $x$, and $  \b_1(0) =   \b_2(0)$.


\section{Adiabatic approximation near $x=0$ for $|\nu| \gg 1$. }\label{app:outer-lift}

 Here we compare the asymptotics of the adiabatic approximation $\hPs$ in terms of the perturbed operator $\hK$ for $|\tau| \to \infty$ with the adiabatic approximation $\Ps$ in terms of the original operator $\K$. We carry out this comparison in the case where
adiabatics can be applied  for $\hPs$.

\vskip5mm

According to \Ref{hWKB1-norm} the principal approximation of the adiabatic mode for the perturbed problem is expressed through perturbed eigenfunctions $\vvph_1(x)$, perturbed eigenvalues $\hb_1(x)$, and the imaginary part of the conversion coefficient $\hat S_{\brr11}(x)$. We may find these quantities near $x=0$ by means of asymptotic expansions from Section \ref{sec:EVP-close}.

 According to \Ref{EF}  and with account of the fact $\sgn N_1^{(0)}=\sgn N_2^{(0)}$ and  \Ref{al-s},  the normalized perturbed eigenfunction reads
\be\label{pertub-eig}
\frac{\vvph_j(x)}{|\hat N_j(x)|^{1/2}} = e^{i  {\delta_j}}
\frac{\a_{j1}^{(0)}(\tau) \vph_1(0) + \a_{j2}^{(0)}(\tau) \vph_2(0)}{\,|\,\, N_1^{(0)}|\a_{j1}^{(0)}(\tau)|^2 +  N_2^{(0)} |\a_{j2}^{(0)}(\tau)|^2 \,\, |^{1/2}}.
\ee
where the $\delta_j$ are arbitrary constants.
To obtain the asymptotics of \Ref{pertub-eig}, we note that from \Ref{al-s} it follows that
\begin{eqnarray}\label{a-asj1+}
&\a_{12}^{(0)} \=_{\tau \to + \infty}	\frac{i\nu}{\tau}
			+ O(\t^{-2}),\qquad
&\a_{12}^{(0)} \=_{\tau \to - \infty}
		2 Q \t + O(1), \qquad \\
&\a_{22}^{(0)} \=_{\tau \to + \infty}
		2 Q \t + O(1),\qquad
&\a_{22}^{(0)} \=_{\tau \to - \infty}
    \frac{i\nu}{\tau}
		+ O(\t^{-2}).
\label{a-asj1-}
\end{eqnarray}
From \Ref{al-s} in view of  \Ref{Thet-chos} we have
\be\label{a-asj2}
		\a_{j1}^{(0)} \=_{\tau\to \pm \infty} \frac{\Bcn0_{12}}{N_1^{(0)} }, \quad j=1,2, \quad \a_{j1}^{(0)}=|\a_{j1}^{(0)}| e^{2i \vartheta_1},\quad \vartheta_2 = - \vartheta_1.
\ee

Formula \Ref{hWKB1-norm} gives the adiabatic approximation.  We note that $\Im{\hat S_{11}}=0$.
Applying the asymptotic formula for eigenvalues as $\t \to \infty$ and the definition of $\nu$ \Ref{nu_gen}, we get
\be
\begin{aligned}
&\frac{\vvph_1 (\t)}{|\hat N_1|^{1/2}} \to
		e^{i  {\delta_1}}\left\{\begin{array}{rl}
		- \frac{\vph_2(0)}{|N_2^{(0)}|^{1/2}}& \qquad {\rm as} \quad\tau\to-\infty,\\
		e^{2i\vartheta_1} \frac{\vph_1(0)}{|N_1^{(0)}|^{1/2}}&\qquad{\rm as} \quad \tau\to +\infty,
		\end{array}
		 \right.\qquad \\
&\frac{\vvph_2 (\t)}{|\hat N_2|^{1/2}} \to
		e^{i  {\delta_2}}\left\{\begin{array}{rl}
		e^{2i\vartheta_1}\frac{\vph_1(0)}{|N_1^{(0)}|^{1/2}}&\qquad {\rm as} \quad \tau\to-\infty,\\
		\, \frac{\vph_2(0)}{|N_2^{(0)}|^{1/2}} & \qquad {\rm as} \quad\tau\to +\infty.
		\end{array}
\right.
\label{vvph_lim}
\end{aligned}
\ee
 We  choose $\delta_1=\delta_2=e^{-i\vartheta_1}$.  Taking into account \Ref{hb-b-t+},\Ref{hb-b-t-}, the definition of canonical mode \Ref{Psi-can}, and the fact that $\Im{\hat S_{jj}}=0$, we find
\begin{eqnarray}
&\hPs_1 = - \Ps_{2-}^{(1)} + o(1), \quad &\hPs_2 =  \Ps_{1-}^{(1)} + o(1) \quad {\rm as} \quad \tau \to -\infty, \label{pert-unpert1} \\
&\hPs_1 = \Ps_{1+}^{(1)} + o(1), \quad &\hPs_2 =  \Ps_{2+}^{(1)} + o(1) \quad {\rm as} \quad \tau \to +\infty, \label{pert-unpert2}
\end{eqnarray}
 which are the required relations between the canonical modes $\hPs_j$ and $\Ps_j,$ $j=1,2$.
\vskip 12pt

We intend to show now that the adiabatic approach works even near $x=0$ if $|\nu|\sim \h^{-\alpha},$ $\quad \alpha>0$. We calculate the main singularity in each approximation by using explicit formulas for an adiabatic approximation for the perturbed $\hK$.

To check that the conditions of applicability of adiabatic approximation are fulfilled, we give below the main order of the next approximations of $\hPhn{n}{1},$ $\hvtn{n}{1}$, which can be found by analogy with the adiabatic expansion of $\Ps_j,$ $j=1,2$ (see Section \ref{sec:wkb-valid}:
\begin{eqnarray}
\hvtn{n}{1}&=& i \frac{ ( \vvph_1, \G \partial_x  \hPhn{n-1}{1} )   }{  \hat{N}_1}
\sim i \hcn{n-1}{12} \hat S_{12} , \quad n\ge1, \label{pert-thet-n}\\
\hPhn{n}{1}&=& \hcn{n}{12} \vvph_2 + \vvph_{1\perp}^{(n)}, \qquad \hcn{1}{12}= i \frac{ \hat S_{\brr21}}{( \hb_1- \hb_2)}, \label{pert-hPhn}\\
\cn{n}{12} &=&
	=\frac{- \hat{N}_2 \sum_{i=1}^{n-1}\hvtn{i}{1} \hcn{n-i}{12}
		+ i ( \vvph_2,\G \partial_x \hPhn{n-2}{1} ) }{ ( \hb_1- \hb_2) \hat{N}_2}, \quad  n\ge 2.			 \label{pert-cn}
\end{eqnarray}
Here $\vvph_{1\perp}^{(n)}$ is $\G$-orthogonal to $\vvph_j$, $j=1,2$. It may have only a lower order singularity near $x=0$ comparing with $\hcn{n}{12}$.

 Formula \Ref{eig-root} and the definition of $\nu$ \Ref{nu_gen} show that the minimal value of eigenvalues in the case of the avoided crossing and for fixed $Q$ can be interpreted in terms of $\nu$:

\be
	 \min(\hb_2-\hb_1) = 2 \sh \sqrt {2 Q |\nu| } + O( \h).
\ee
Applying it to \Ref{pert-thet-n}--\Ref{pert-cn}
we show that $|\hat S_{12}|=|\hat S_{21}|=Q |B_{12}|/(2 \h |\hb_1-\hb_2|^2),$
$|c_{12}^{(1)}| \le 1/(8 \h |\nu|).$ Considering the formulas \Ref{pert-thet-n} and \Ref{pert-cn} step by step, we deduce that $|\hvtn{n}{1}/2(\hb_1-\hb_2)| \sim 1/(8 \h |\nu|)^n$, $|c_{12}^{(n)}| \sim 1/(8 \h |\nu|)^n.$


\section{ The properties of the transition matrix}\label{app:T_prop}
The general properties of the transition matrix follow from the flux conservation law \Ref{conser}.
%
For two solutions $\Ps_j$, $j=1,2$ having $\Ps_{j-}^{(0)}$ as their asymptotic on the left of the degeneracy point
\be
	\Ps_j \seq_{ {x \ll -\sh}}  \Ps_{j-}^{(0)} +O(\sh),
	\label{PS-}
\ee
their asymptotics on its right can be derived via \Ref{trMa} and will be given by
\be
	\Ps_j \seq_{ {x \gg \sh}} t_{j1}  \Ps_{1+}^{(0)}+t_{j2} \Ps_{2+}^{(0)}+O(\sh),
	\label{PS+}
\ee
where $t_{jk}$, $j,k=1,2,$ are the $\T$ matrix entries. At the same time, the relation
\be
	(\Ps_j, \G \Ps_k) = const , \quad
	j,k=1,2
	\label{const}
\ee
must hold and the constants are the same on both sides of the degeneracy point.

Calculating  the scalar product in \Ref{const} for all values of the indices by using \Ref{PS-} and \Ref{PS+} for both sides of the degeneracy point, and equating the results, we obtain
\begin{eqnarray}
   N_1  = |t_{11}|^2   N_1 + |t_{21}|^2   N_2, \label{T-1}\\
   N_2  = |t_{12}|^2   N_1 + |t_{22}|^2   N_2, \label{T-2}\\
0 = \overline{t_{11}} t_{12}   N_1 + \overline{t_{21}} t_{22}   N_2.  \label{T-3}
\end{eqnarray}
Equation \Ref{T-3} yields
\be\label{not-kappa}
\frac{\overline{t_{11}}}{t_{22}} = - \frac{  N_2}{   N_1} \frac{\overline{t_{21}}}{t_{12}} \equiv \gamma,
\ee
where we introduce the notation $\gamma$. The last formula enables us to derive from \Ref{T-1} and \Ref{T-2} that
$\det{\T} = {1}/{ \gamma} = \overline{\gamma}.$  Therefore $|\gamma|=1$, i.e., $\gamma$ is a phase factor.
We can govern $\gamma$ by changing lower limits of integration in adiabatic formulas.
Then we deduce that for an appropriate choice of the arbitrary phase of the adiabatic mode, the matrix $\T$ has the following properties:
\be\label{gen-prop-T}
{\det{\T}} =1, \quad
\T = \(\begin{array}{cc}
		 t_{11} &  t_{12}\phantom{\Big|} \\
	- \overline{t_{12}} \frac{  N_1}{  N_2}\
		& \overline{t_{11}}\phantom{\Big|}
		\end{array}\).
\ee
If ${\rm sgn}{  N_1}={\rm sgn}{  N_1}$, the matrix $\T$ is unitary $\T \T^+= \I,$ where $\I$ is the identity matrix.



\section{Examples} \label{app:examples}

We start our list of examples with a well-known case of real turning points in the  stationary Schroedinger equation, which was probably the first ever studied problem. Although it served for the authors as an incentive for the studies, this example needs an alteration of the method presented in the paper due to the presence of the Jordan block in the operator $\K$ at the degeneracy point.

Then we elaborate in full detail the application of our method to the description of the electrons scattering in graphene in the external potential. After formulating the problem in our terminology and checking the applicability of our results to this physical problem (following the first two steps of the Recipe, Sec. \ref{sec:recipe}), the only calculation needed to obtain the transition matrix is to substitute the values of parameters describing the degeneracy point into the formula for the transition matrix \Ref{T}.


\subsection{Stationary Schroedinger equation}

The stationary Schroedinger equation in the one-dimensional space with external potential $U=U(x)$ is written as \cite{qm_landavshitz_59}
\be\label{stat-Sch}
	-\frac{\hbar^2}{2m}\psi''(x)+ (U(x)-E)\psi(x)=0.
\ee
Introducing a new two-component unknown function
\be
	\Ps=(\psi(x), -i \h\, \psi'(x) )^T,
\ee
we rewrite the original equation \Ref{stat-Sch} in the form
\be
	\K\Ps=-i\h \G\frac{\partial\Ps}{\partial x}
\ee
with Hermitian operators
\be
  \K\equiv\left(
           \begin{array}{cc}
            2m (E-U(x))\  & 0\\
             0 & 1\\
           \end{array}
         \right),\qquad
\G=      \left(
           \begin{array}{cc}
             0\ &1\\
             1\ &0 \\
           \end{array}
         \right).
\ee
The conversation law \Ref{conser} in this case has the meaning of the probability current
\be
	(\Ps,\G\Ps)=i\h \( \psi(x)  \frac{d\psi^*(x)}{dx}-\psi^*(x)  \frac{d\psi(x)}{dx}\),
\ee
which  is constant for the static wave functions.

The eigenvalue problem  $\K \vph= \b\G \vph$ gives
\be
	 \b_j=(-1)^{j+1} \sqrt{2m(E-U(x))}, \qquad
		 \vph_j=(1, \b_j)^T, \quad j=1,2.
\ee
At the degeneracy points $x=\varkappa$, where
\be
	\b_1(\varkappa)=\b_2(\varkappa)=0,
\ee
the eigenfunctions become  linear dependent
\be
	\vph_1(\varkappa)=\vph_2 (\varkappa)= (1,0)^T,
\ee
thus contradicting our assumption 2 in Section \ref{sec:stat-pr} -- the operator contains a Jordan block at the degeneracy point.

We draw a conclusion that although this problem can be treated using the method developed in this paper, the  result obtained for the transition matrix \Ref{T} is not applicable. Appropriate modifications are the subject of future works.

\subsection{Dirac equation in $2+1$ dimensions}
\label{app:dirac}
Properties of the electronic excitations in graphene are described by the massless Dirac equation in $(2+1)$ - dimensional space \cite{Katsnelson}. In the presence of the external potential $U=U(x,y)$, the stationary wave function $e^{iEt}\wt{\Ps}(x,y)$ must satisfy
\be
	\(v_F {\bm \sigma}\cdot \hat{\bf p} +U\)\wt\Ps(x,y)=E\wt\Ps(x,y),
\label{dirac}
\ee
where
${\bm \sigma}=(\sigma_x,\sigma_y)$ is a pair of Pauli $2\times 2$ matrices,
$\hat{\bf p}=-i\h \nabla$; for simplicity we put the Fermi velocity equal to one, $v_F=1$.

One of the physical problems described by \Ref{dirac} is the electron scattering on an electrostatic potential barrier inside graphene. If the external potential depends on one variable only, $U=U(x)$, we can separate the $x$-derivative in the equation \Ref{dirac}.

Following our Recipe, we proceed to the Fourier transform,
\be\nonumber
{\Ps}(x,p_y)=\int\limits_{\mathbb{R}} d y e^{-ip_y y/\h}\wt{\Ps}(x,y)
\ee
to obtain
the equation in the form \Ref{newSch}
\be
	\hat{\K}\Ps=-i\h \G\frac{\partial\Ps}{\partial x}
\ee
with Hermitian operators
\be
  \hat{\K}\equiv\left(
           \begin{array}{cc}
             E-U(x) & -i p_y \\
             i p_y & E-U(x)\\
           \end{array}
         \right),\qquad
\G=      \left(
           \begin{array}{cc}
             0&1\\
             1&0 \\
           \end{array}
         \right).
\ee
The $\G$-eigenvalues of $\hat\K$ are
\be
	\hb_{j}=(-1)^{j+1}  \sqrt{(E-U)^2-p_y^2}, \quad j=1,2,
\ee
with eigenfunctions
\be
\vvph_{j}=  \left(
           \begin{array}{c}
            E-U(x) \\
             \hb_{j}+ip_y \\
           \end{array}
         \right).
\ee
Their normalization is
\be
\hat N_{j}= 2 (E-U) \Re \hb_{j}.
\label{graN}
\ee
The conservation law \Ref{conser} in this case corresponds to the conservation of the $x$-component of the electron current $j^\mu$ \cite{Bogoliubov}, due to the simple fact that $\G=\sigma_x$,
\be
	(\Ps,\G\Ps) =
	\Ps^\dag \sigma_x\Ps \equiv  j_x,
\ee
and can naturally be both positive and negative as  is obvious from \Ref{graN}.

Depending on the form of the potential and relative values of $E$, $p_y$,  different types of (quasi)degeneracy of the eigenvalues, $\hb_1(\varkappa)\seq\hb_2(\varkappa)$ may exist.
The most studied case \cite{koen,zalip} is when the electrons are incident almost perpendicularly to the potential barrier. The interest in this case was invoked by the presence of the so-called Klein paradox \cite{klein}, which is characterized by the unit probability of tunnelling through the barrier, for applications in grahene see \cite{klein-gra}.

The scattering of electrons on a potential close to normal   is characterized  by small values of $p_y$.
In particular, if we assume that  $p_y$  is of order $\sh$, $p_y=\sh\, p$, $p=O(1)$, we can separate the perturbation $\B$ from the original operator in the following way:
\be
	\hat{\K}=\K+\sh \B,\ee
where
\be
	      \K = \left(
           \begin{array}{cc}
             E-U(x) & 0 \\
             0 & E-U(x) \\
           \end{array}
         \right),\qquad
	      \B = \left(
           \begin{array}{cc}
             0& -i p \\
             ip &0\\
           \end{array}
         \right).
\label{H0B gra}
\ee
This completes the first stage of the Recipe (see Section \ref{sec:recipe}).

On the second stage, one should solve the eigenvalue problem for the original operator pencil
$\K  \vph= \b\G \vph$. It is straightforward in this case that
\be
	 \b_j =(-1)^j (E-U(x)), \quad
 \vph_j= \frac1{\sqrt2}\left(
           \begin{array}{c}
             1 \\
             (-1)^j \\
           \end{array}
         \right),\qquad j=1,2,
\label{m vph gra}
\ee
the sign of normalization reflects the electric charge of these modes. So, in this case we have
\be
	N_1 N_2 = -1.
\ee

The degeneracy points $x=\varkappa$ are those where $\b(\varkappa)=0$, i.e.,
\be
	E= U(\varkappa).
\ee
Supposing that on the interval of interest there is only one such point $\varkappa=0$, we conclude that all the assumptions of the Section \ref{sec:stat-pr} and \Ref{11}, in particular, are satisfied if the potential has a nonvanishing first derivative at  $x=\varkappa$, $Q\equiv\frac{\partial U}{\partial x}(0)\ne 0$.

For constructing the canonical modes on the third stage of the Recipe, we calculate the matrix entries
\be
	\Bc_{12}=-\Bc_{21}=-ip, \qquad
	\Bc_{11}=\Bc_{22}=0
\ee
and other necessary ingredients
\be
	S_{ij}=0,\qquad
	Q=U'(0),\qquad
	\nu=- i \frac{p^2 }{2 Q},
\ee
\be
	\theta_a=\arg \Bc_{12}+\pi/2=0, \qquad
	b=0,
\ee
\be
	\varkappa_\pm= \pm \sh \, p/Q.
\ee
We also note that in this case the eigenvalues are symmetric, $\b_1=-\b_2$, so that
\be
\b_0=\b_{av}^{(1)}=0.
\ee

The canonical modes are
\be
 {\Ps}_{j\pm}^{(0)} =   \vph_j (x)
  e^{i/\h \,\int\nolimits^x_{x^{}_{\pm}} \hb_j^{pr}(x',\h) dx' },
 	\label{Psi-can-ell}
\ee
\be
  \hb_1^{pr}(x',\h) = \left\{
  \begin{array}{ll}
	(E-U(x'))-\h\frac{  p^2}{ 2 (E-U(x')) } \
		& \qquad{\rm for\ } |x'|\in (x^*, |x| ), \\
 - Q \,{\rm sgn}x' \sqrt {x'^2 -  \h p^2/Q^2 }
		  & \qquad {\rm for\ }  |x'|\in (0,x^*).\label{hbet-pr-gra}
\end{array}
\right.
  \ee
and $\hb_2^{pr}(x',\h)=-\hb_1^{pr}(x',\h)$. The transition matrix is given by \Ref{T} with $\nu$ from above.

Thus, our method can be applied readily to the description of the nearly normal scattering of electrons on an external potential. This case was investigated successfully in \cite{koen,zalip} by a different method, and it is a straightforward task to verify that our result for the transition matrix is in agreement with their results.

The more interesting case where $U(x) = E+O(x-\varkappa)^2$, i.e., where the energy of the incoming electrons are equal (or close) to the extremal value of the potential is the theme of future work.

\subsection{ Wave equation}
Considering the monochromatic solutions $u = e^{- i \omega t} v$ of the wave equation
with a nonconstant velocity
\begin{equation}
	\frac{1}{c^2} {{\partial^2 u} \over {\partial t^2}} - \Delta u = 0,
\end{equation}
we obtain a Helmholtz equation for $v$
\be\label{Helmh}
	\Delta v + \frac{\omega^2}{c^2} v =0.
\ee

Suppose  the velocity of  wave propagation depends only on one spatial variable  $c=c(x'/L)$, where $L$ is the characteristic scale of the velocity variations. A solution $v$  dependent on $x'$ only, will satisfy then a second order ordinary differential equation
\begin{equation}
	{d ^2 v \over d x'^2} +  \frac{\omega^2}{c^2( {x'\over L})} v = 0.
\label{second-undim}
\end{equation}
To write this second order equation in the form of a first order system like \Ref{newSch}, one may take the first derivative of $v$ as a new variable, i.e., putting
\be\label{var-new}
	\Psi_1 = v,\quad
	\Psi_2 = - \frac{i}{ k_0} {dv \over d x'}, \qquad k_0 = \frac{\omega}{c_0},
\ee
where $c_0=c(0)$, and we assume $c(x'/L)\sim c_0$.
Then equation \Ref{second-undim} can be written as
\begin{equation}
  k_0 \left(
             \begin{array}{cc}
               c_0^2\over {c^2({x'/ L})} & 0 \\
                   0 &  1 \\
             \end{array}
           \right)  \left(
             \begin{array}{c}
               \Psi_1 \\
               \Psi_2 \\
             \end{array}
           \right)
          = - i   \left(
             \begin{array}{cc}
               0 & 1 \\
               1 & 0 \\
             \end{array}
           \right){\partial \over {\partial x'}}
           \left(
             \begin{array}{c}
               \Psi_1 \\
               \Psi_2 \\
             \end{array}
           \right).
\label{second-matr2}
\end{equation}
Upon introducing a dimensionless variable $x = x'/L$ and dividing (\ref{second-matr2}) by $k_0$, we rewrite it finally in the form \Ref{newSch},
where
\begin{equation}
\K  =   \left(
             \begin{array}{cc}
               c_0^2\over {c^2({x})} & 0 \\
                   0 & 1 \\
             \end{array}
           \right),\quad
       \G =  \left(
             \begin{array}{cc}
               0 & 1 \\
               1 & 0 \\
             \end{array}
           \right),
\end{equation}
and the (small) parameter is $\h = 1/(k_0 L)$. Thus, the condition $\h\ll 1$ implies that  $k_0 L \gg 1$.

The eigenvalue problem \Ref{newEVP} for this case has the following solution:
\be
	\b_j(x)=(-1)^{j}  c_0 / c({x}), \qquad
	\vph_j=(1,\b_j).
\ee
The degeneracy points, where $\b_1=\b_2$, correspond to $c(x)\to\infty$. In the electromagnetic media, for instance, the light velocity is inverse proportional to the  square root of dielectric and magnetic permeabilities. So, at the points where one of them vanishes (as happens in plasma \cite{Budden72}), $c(x)$ tends to infinity.

The Helmhotlz equation with many spatial coordinates can be considered analogously. For instance, in the case of a two-dimensional waveguide
\begin{equation}
	{\partial^2 u \over \partial x'^2} + {\partial^2 u \over \partial z'^2} +
	  \frac{\omega^2}{c^2({ x'\over L}, {z'\over H})} u = 0,
	\label{two-dim-Helm}
\end{equation}
with $u$ defined inside a strip with some boundary conditions, e.g.,
\begin{equation}\label{bound}
	u{|_{z'=0}} = 0, \qquad \left.\frac{d u}{dz'}\right|_{z'=H} = 0,
\end{equation}
we can introduce once again the new variables \Ref{var-new} to obtain
\begin{equation}
  k_0 \left(
             \begin{array}{cc}
            {1\over k_0^2} {d^2  \over d z'^2} + \frac{c_0^2} {c^2({x'\over L},{z' \over H})} & 0 \\
                   0 &  1 \\
             \end{array}
           \right)  \left(
             \begin{array}{c}
               \Psi_1 \\
               \Psi_2 \\
             \end{array}
           \right)
          = - i   \left(
             \begin{array}{cc}
               0 & 1 \\
               1 & 0 \\
             \end{array}
           \right){\partial \over {\partial x'}}
           \left(
             \begin{array}{c}
               \Psi_1 \\
               \Psi_2 \\
             \end{array}
           \right).
\label{second-matr3}
\end{equation}
In terms of dimensionless $x = x'/L$, $z=z'/H$, this equation takes the form \Ref{newSch} with $\h = 1/(k_0 L),$ and
\begin{equation}
\K  =   \left(
             \begin{array}{cc}
             {1\over (k_0 H)^2} {d^2  \over d z^2} +  \frac{c_0^2} {c^2(x, z)} & 0 \\
                   0 & 1 \\
             \end{array}
           \right),\quad
       \G =  \left(
             \begin{array}{cc}
               0 & 1 \\
               1 & 0 \\
             \end{array}
           \right).
\end{equation}

The conservation law (\ref{conser}) for (\ref{second-matr2},\ref{second-matr3}) is given by
\be
	(\Ps, \G \Ps) = \frac2{c_0} \Re\( -i \omega \bar{u} {d u \over d x} \).
\ee
For the original time-dependent solution $u$ it has the meaning of the time-averaged flux of energy in the $x$ direction.


\end{document}